\documentclass[%
 reprint,
 amsmath,amssymb,
 aps,
]{revtex4-2}

\usepackage{amsthm}
\usepackage{graphicx}% Include figure files
\usepackage{dcolumn}% Align table columns on decimal point
\usepackage{bm}% bold math
\usepackage{xcolor}
\usepackage[bottom]{footmisc}
\usepackage{fancyhdr}

\newtheorem{definition}{Definition}
\newtheorem{example}{Example}
\newtheorem{theorem}{Theorem}
\newtheorem{lemma}{Lemma}
\newtheorem{proposition}{Proposition}
\newtheorem{remark}{Remark}
\newtheorem{corollary}{Corollary}

\begin{document}

\preprint{APS/123-QED}

\title{Criteria of absolutely separability from spectrum for qudit-qudits states}% Force line breaks with \\
%\thanks{A footnote to the article title}%

\author{Liang Xiong}
 \altaffiliation{liang.xiong@polyu.edu.hk}%Lines break automatically or can be forced with \\
\author{Nung-sing Sze}%
 \email{raymond.sze@polyu.edu.hk (Corresponding author)}
\affiliation{Department of Applied Mathematics, The Hong Kong Polytechnic University, Hung Hom, Hong Kong, China
 %This line break forced with \textbackslash\textbackslash
}%

%\collaboration{MUSO Collaboration}%\noaffiliation

%\author{Charlie Author}
% \homepage{http://www.Second.institution.edu/~Charlie.Author}
%\affiliation{
% Second institution and/or address\\
% This line break forced% with \\
%}%
%\affiliation{
% Third institution, the second for Charlie Author
%}%
%\author{Delta Author}
%\affiliation{%
% Authors' institution and/or address\\
% This line break forced with \textbackslash\textbackslash
%}%
%
%\collaboration{CLEO Collaboration}%\noaffiliation

\date{\today}% It is always \today, today,
             %  but any date may be explicitly specified

\begin{abstract}
Separability from the spectrum is a significant and ongoing research topic in quantum entanglement. In this study, we investigate properties related to absolute separability from the spectrum in qudits-qudits states in the bipartite states space $\mathcal{H}_{mn}=\mathcal{H}_m \otimes \mathcal{H}_n$. Firstly, we propose the necessary and sufficient conditions for absolute separable states in the Hilbert space $\mathcal{H}_{4n}$. These conditions are equivalent to the positive semidefiniteness of twelve matrices resulting from the symmetric matricizations of eigenvalues. %However, determining these twelve linear matrix inequalities is a challenging task. Motivated by this, we present sufficient conditions based on the construction of the twelve matrices.
%the fact that all third-order principal submatrices of the twelve matrices are positive semidefinite.
Furthermore, we demonstrate that this sufficient condition can be extended to the general $\mathcal{H}_{mn}$ case, improving existing conclusions in the literature. These sufficient conditions depend only on the first few leading and last few leading eigenvalues, significantly reducing the complexity of determining absolute separable states.
On the other hand, we also introduce additional sufficient conditions for determining that states in $\mathcal{H}_{mn}$ are not absolutely separable. These conditions only depend on $2m-1$ eigenvalues of the mixed states. Our sufficient conditions are not only simple and easy to implement. As applications, we derive distance bounds for eigenvalues and purity bounds for general absolutely separable states.
\begin{description}
\item[Keyword]
Absolutely separability; eigenvalue; positive semidefiniteness; quantum entanglement
%\item[Structure]
%You may use the \texttt{description} environment to structure your abstract;
%use the optional argument of the \verb+\item+ command to give the category of each item.
\end{description}
\end{abstract}

%\keywords{Suggested keywords}%Use showkeys class option if keyword
                              %display desired
\maketitle
%\begin{keyword}
%Absolutely separability; eigenvalue; positive semidefiniteness; quantum entanglement
%\end{keyword}

%\tableofcontents

\section{Introduction}

The phenomenon of entanglement in composite systems is a fundamental and significant feature in quantum information science \cite{quantuminformation, PPT2, GUHNE2009}. However, quantifying entanglement has proven to be difficult. Determining whether a quantum state is entangled or not is essential in many practical applications. Despite the great interest in entangled quantum states, their complete characterization remains an unsolved problem. Nevertheless, there are entanglement criteria such as the positive partial transposition (PPT) criterion \cite{PPT1, PPT2, 2016PPT}, Bell inequalities \cite{Bell, Bell2, Bell3}, entanglement witnesses \cite{biEW2009, MEW2013, NEW2018, norm1, Sarbicki_2008}, and the computable cross norm or realignment (CCNR) criterion \cite{Li_2011, CCNR}. The PPT criterion gives a complete characterization of entanglement in the qubit-qubit and qubit-qutrit cases.
%Hilbert spaces of dimension $2\times 2$ and $2\times 3$
\cite{10.1007/BF02391860, WORONOWICZ1976165, HORODECKI1996}.

For a mixed state $\rho$ on a Hilbert space $\mathcal{H}_N$ of dimension $N$, what are the conditions on the spectrum of $\rho$ that guarantee $\rho$ is separable with respect to any decomposition of $\mathcal{H}_{N}$ as a tensor product $\mathcal{H}_{n} \otimes \mathcal{H}_{m}$ of two Hilbert spaces of dimensions $n$ and $m$ respectively with $N = nm$? This problem, listed as problem 15 of open problems in Quantum Information Theory on the website of the Institute of Mathematical Physics at the TU Braunschweig \cite{krueger2005open}, seeks to characterize the spectrum conditions that determine separable states. These states remain separable under any global unitary transformation and cannot become entangled. This is crucial for input states that rely on entanglement to achieve related operations. Solving this problem provides insight into the nature of entanglement and the structure of quantum states. However, specifying a quantum state acting on an $N$-dimensional space requires $N^2-1$ real parameters, making it challenging to completely reconstruct the state via tomography in experiments \cite{tomography, tomography2, tomography3}. Therefore, determining separability from the eigenvalues offers a simple experimental approach to determine separability for some states.

Despite the challenges in solving the separability problem from the spectrum, progress in this area has enhanced our understanding of the structure of quantum states and the nature of entanglement. Previous observations have shown the existence of a ball of separable states centered at the maximally mixed state $\frac{1}{mn}(I_{m}\otimes I_{n})$, where every state within this ball is separable \cite{Ball02, Ball98, Ball05}. The exact Euclidean radius of the largest such ball is known, and any state within this ball is separable. However, there are also absolutely separable states outside of this ball \cite{Vidal1999}.

Absolute separable states \cite{AbsolutelySep2001,AbsolutelySep2007,AbsolutelySep2013} are a specific type of separable state that remains separable under any global unitary transformation $U$, ie., $U\rho U^\dag$ is always separable. These quantum states can be determined to be separable based solely on their eigenvalues. The characterization for separability from the spectrum was presented in the case of the qubit-qubit system \cite{As201}. It was shown that if $\rho\in \mathcal{H}_{2} \otimes\mathcal{H}_{2}$ has eigenvalues $\lambda_1 \ge \lambda_2 \ge \lambda_3 \ge \lambda_4$ and satisfies the following inequality:
\begin{equation*}
  \lambda_{1}\leq \lambda_{3}+2\sqrt{\lambda_{2}\lambda_{4}},
\end{equation*}
then it is absolutely separable.

Similarly, there are absolutely PPT states, which are states that remain positive partial transpose (PPT) under any global unitary transformation \cite{Ball98,Johnston15}. For qutrit-qudit systems ($\mathcal{H}_{3} \otimes\mathcal{H}_{n}$), necessary and sufficient conditions on the spectrum of absolutely separable states have been established, which involve a set of linear matrix inequalities on the eigenvalues of $\rho$ \cite{AbsolutelySep2007}.

Moreover, surveys conducted by Johnston \cite{AbsolutelySep2013} have revealed that in the case of qudit-qubit systems ($\mathcal{H}_{2} \otimes\mathcal{H}_{n}$), the set of absolutely PPT states coincides with the set of absolutely separable states. This means that a necessary and sufficient criterion for absolute separability exists when one of the subsystems is a qubit system. Specifically, if the eigenvalues of the state $\lambda_1 \ge \cdots \ge \lambda_{2n}$ (in decreasing order) satisfy the following inequality,
\begin{equation}\label{ieqAS13}
  \lambda_{1} \le \lambda_{2n-1}+2\sqrt{\lambda_{2n-2}\lambda_{2n}},
\end{equation}
then the qubit-qudit state is absolutely separable.

It is well established from various studies that the set of absolutely separable bipartite states forms a convex and compact set in any arbitrary dimensional Hilbert space \cite{PRA2014}. Additionally, Halder et.\,al. \cite{AS21} has characterized the boundary of the convex compact set of absolutely separable states in the qubit-qudit system. They discuss the properties of boundary points and extreme points of absolutely separable states, including those inside and outside the maximal ball, based on different matrix ranks. Furthermore, they also consider the reverse process, exploring the range of noise parameters that can produce absolutely separable states from entangled states when the states send through local noisy channels.

On the other hand, the Jivulescu et.\,al. \cite{JIVULESCU2015276} introduces the set of absolutely RED states, which remain positive under the reduction map applied to the second subsystem after any global unitary transformation. They provide necessary and sufficient conditions on the spectrum in the form of a family of linear inequalities. The entanglement criteria for spectrum and bounds for absolutely RED are established in this literature \cite{JIVULESCU2015276}. For instance, if $A$ is a self-adjoint positive semidefinite operator on the Hilbert space $\mathcal{H}_{mn}=\mathcal{H}_m \otimes \mathcal{H}_n$, and $\lambda_1, \ldots, \lambda_{m n}$ are the eigenvalues of $A$ in decreasing order, then $A$ is absolutely PPT if the following inequality holds:
\begin{equation}\label{mneq1}
  \lambda_{mn}+\lambda_{mn-1}+\lambda_{mn-2}\geq \lambda_1.
\end{equation}

The problem of absolute separability has analogs in various settings, such as continuous variable systems \cite{variable1,variable2}, quantum channel theory \cite{Filippov2017}, symmetric systems \cite{CHAMPAGNE2022}, and quantum coherence \cite{Acoherent22,PhysRevA.95.012318}. In the context of quantum channel theory, the authors \cite{Filippov2017} have explored the properties of absolutely separable states and established an upper bound on the purity of absolutely separable states, particularly in tripartite systems. They have also introduced the class of absolutely separating maps and presented their fundamental properties. Additionally, specific quantum channels are considered within families of quantum maps, such as local depolarization of qubits, local unital maps on qubits, and generalized Pauli diagonal channels constant on axes. In a similar vein to the absolute separability problem in quantum resource theory, Johonston et.\,al. \cite{Acoherent22} derive several necessary and sufficient conditions for absolutely $k$-incoherent states, which depend solely on their eigenvalues and result in hyperbolicity cones associated with elementary symmetric polynomials.

The absolute separability problem has been addressed in symmetric separable states, where the separability of symmetric quantum states is guaranteed only by their spectra \cite{CHAMPAGNE2022}. The authors in \cite{CHAMPAGNE2022} demonstrate that, in the case of $m=2$ or $m=3$, a symmetric quantum state $\rho$ in $\mathcal{H}_m \otimes \mathcal{H}_m$ is absolutely symmetric separable if and only if every symmetric matricization of the eigenvalues of $\rho$ is positive semidefinite.  It is evident that only one specific symmetric matricization needs to be positive semidefinite to ensure that all matricizations are positive semidefinite in the case of $m=2$ and $m=3$.

However, the situation becomes substantially more complicated in the case of $m\ge 4$, since many more symmetric matricization will be involved in the study. This is still an open problem in the field. Therefore, further exploration of absolute separability in the basis of eigenvalues of the density matrix in the case of $m\ge 4$ is of interest.

In this paper, we explore the properties of absolutely separable states based on the spectrum for qudit-qudit states. Firstly, we establish necessary and sufficient conditions for absolutely separable states in $\mathcal{H}_{4 n}=\mathcal{H}_4 \otimes \mathcal{H}_n (n\geq 4)$. We showed that the absolutely separability condition is equivalent to the positive semidefiniteness for twelve symmetric matricizations of the $4n$ eigenvalues. However, determining their positive semidefiniteness is a challenging task. To address this, we introduce some sufficient conditions for absolutely separable states and non-absolutely separable states in $\mathcal{H}_{4 n}=\mathcal{H}_4 \otimes \mathcal{H}_n (n\geq 4)$, based on the fact that all third-order principal submatrices of the twelve aforementioned matrices are positive semidefinite. Building on this, we aim to establish sufficient conditions for the general $\mathcal{H}_{mn}=\mathcal{H}_m \otimes \mathcal{H}_n$ case, which depend solely on the first few leading and last few leading eigenvalues of the states and are easy to implement. Furthermore, we demonstrate that our findings overperform those existing results in low dimensions ($n\geq m=2,3$).  Also the condition only relies on an inequality of $2m-1$ eigenvalues to avoid the NP-hard problem of positive semidefiniteness of large-scale parameter matrices. Additionally, we introduce some sufficient conditions that states in $\mathcal{H}_{mn}=\mathcal{H}_m \otimes \mathcal{H}_n$ are not absolutely separable, which depend only on $2m-1$ eigenvalues of states. We also investigate the distance bounds of eigenvalues and the bounds of purity for general absolutely separable states in $\mathcal{H}_{mn}$ based on our conditions.

This paper is organized as follows: In section 2, we provide some preliminary knowledge in mathematics and quantum entanglement related to this study. In section 3, we present the necessary and sufficient conditions of absolutely separable states for $\mathcal{H}_{4 n}=\mathcal{H}_4 \otimes \mathcal{H}_n (n\geq 4)$, and introduce some easily achievable sufficient conditions. We also consider the sufficient conditions for states that are not absolutely separable in $\mathcal{H}_{4 n}$. In section 4, we demonstrate the sufficient conditions of absolutely separable states based on the spectrum for the general $\mathcal{H}_{mn}=\mathcal{H}_m \otimes \mathcal{H}_n (m\le n)$ case, and our results improve some existing results in literature. Additionally, we present some sufficient conditions for states that are not absolutely separable based on the spectrum. Moreover, we also explore the distance bounds of eigenvalues and the bounds of purity for general absolutely separable states in $\mathcal{H}_{mn}$ based on our conditions.

\section{Preliminaries}

%\subsection{Preliminaries of mathematics}

In the subsequent sections of this paper, we use the symbol $\mathbb{N}_{+}$ to denote the set of positive integers, and $\mathbb{C}$ and $\mathbb{R}$ to denote the sets of complex numbers and real numbers, respectively. We represent $n$-dimensional complex (real) vectors as $\mathbb{C}^{n}$ ($\mathbb{R}^{n}$), and complex (real) $n\times n$ matrices as $M_n(\mathbb{C})$ ($M_n(\mathbb{R}))$. Additionally, we denote the group of permutations of the set $\{1, \ldots, n\}$ as $S_n$. A square matrix $A$ of dimension $n$ is considered positive semidefinite (PSD) if $x^{\top}Ax\geq 0$ for any nonzero vector $x\in \mathbb{R}^{n}$,  and we denote by $A\succeq O$.  Furthermore, a matrix $A=(a_{ij})\in M_n(\mathbb{C})$ is classified as a diagonally dominant matrix if, for all $i\in \{1,\dots,n\}$, the following condition holds:
\begin{equation*}
  |a_{ii}|\geq \sum\limits_{i=1,j\neq i}^{n}|a_{ij}|.
\end{equation*}
If $A$ is a diagonally dominant matrix with positive diagonal elements, then $A$ is also positive semidefinite (PSD). Additionally, for $k \in \mathbb{N}_{+}$, we denote $\mathcal{H}_k$ as a Hilbert space of dimension $k$, $H_{+}(k)$ as the cone of positive semidefinite (PSD) complex Hermitian matrices of size $k \times k$, and $U(k)$ as the group of unitary matrices of size $k \times k$.

In quantum information theory, a pure quantum state $|v\rangle\in \mathbb{C}^n$ is a normal vector, and a mixed state $\rho\in M_n(\mathbb{C})$ is a positive semidefinite (PSD) Hermitian matrix with  $\operatorname{Tr}(\rho)=1$. A mixed state $\rho \in M_m(\mathbb{C}) \otimes M_n(\mathbb{C})$ is called separable if there exist pure states $\{|w_{i}\rangle \}_i\subseteq \mathbb{C}^m$ and $\{|v_{i}\rangle \}_i\subseteq \mathbb{C}^n$ such that
\begin{equation*}
  \rho=\sum\limits_{i}p_{i}|w_{i}\rangle \langle w_{i}| \otimes |v_{i}\rangle \langle v_{i}|,~~ p_{i}\geq 0, ~~\sum p_{i}=1,
\end{equation*}
where $\otimes$ denotes the Kronecker product, and $\langle v_{i}|$ is the dual (row) vector of $|v_{i}\rangle$. In other words, $\rho$ is separable if and only if it can be written in the form
$$\rho=\sum\limits_{i}X_{i}\otimes Y_{i},$$
where $X_{i}\in M_m(\mathbb{C})$ and $Y_{i} \in M_n(\mathbb{C})$ are  positive semidefinite matrices. If the state $\rho$ is not separable, it is called an entangled state.

%{\bf [To Review]}
%The partial transpose of a density matrix $\rho=\sum\limits_j X_{j}\otimes Y_{j} \in M_m(\mathbb{C}) \otimes M_n(\mathbb{C})$ is defined as
%$$ \rho^{\Gamma}:=\sum_j X_{j}\otimes Y_{j}^{\top}.$$
%then $\rho$ being separable implies that $\rho^{\Gamma}$ is positive semidefinite. If a mixed state $\rho$ is such that $\rho^{\Gamma} \succeq O$  then we say that it has a positive partial transpose (PPT).  The set of separable states is a subset of the set of PPT states.

Considering a bipartite state $\rho$ it can be written in some operator basis $\rho=\sum_{i j k l} \rho_{i j k l}|i\rangle\langle j|\otimes| k\rangle\langle l|$. The partial transpose with respect to the second subsystem is defined as
$$ \rho^{\Gamma}=\sum_{i j k l} \rho_{i j k l}|i\rangle\langle j|\otimes(|k\rangle\langle l|)^{\top}=\sum_{i j k l} \rho_{i j k l}| i\rangle\langle j|\otimes| l\rangle\langle k|. $$
%$$
%\begin{aligned}
%& \rho^{T_A}=\sum_{i j k l} \rho_{i j k l}(|i\rangle\langle j|)^T \otimes|k\rangle\langle l|=\sum_{i j k l} \rho_{i j k l}| j\rangle\langle i|\otimes| k\rangle\langle l| \\
%& \rho^{T_B}=\sum_{i j k l} \rho_{i j k l}|i\rangle\langle j|\otimes(|k\rangle\langle l|)^{\top}=\sum_{i j k l} \rho_{i j k l}| i\rangle\langle j|\otimes| l\rangle\langle k| .
%\end{aligned}
%$$

Consequently, a state $\rho$ is positive partial transpose (PPT) means that  $\rho^{\Gamma}$  have only nonnegative eigenvalues.
If a bipartite state $\rho$ is separable, then $\rho^{\Gamma}$ is positive semidefinite.
The set of separable states is a subset of the set of PPT states.

\iffalse
If $W$ is block-positive but not positive semidefinite, it is called an entanglement witness, since it is then the case that $\operatorname{Tr}(W \sigma) \geq 0$ for all separable $\sigma \in M_m(\mathbb{C}) \otimes M_n(\mathbb{C})$, but there exists some (necessarily entangled) mixed state $\rho \in M_m(\mathbb{C}) \otimes M_n(\mathbb{C})$ such that $\operatorname{Tr}(W \rho) < 0$.
\fi

For any $m, n\in \mathbb{N}_{+}$ and $p=\min \{m,n\}$, let $p_{+}=\frac{p(p+1)}{2}$, $p_{-}=\frac{p(p-1)}{2}$, and let $S_{+}=\{(k, l) \mid 1 \leqslant k \leqslant l \leqslant p\}$ and $S_{-}=\{(k, l) \mid 1 \leqslant k<l \leqslant p\}$ be two sets of index pairs with cardinalities $p_{+}, p_{-}$, respectively.
Let $x = (x_1, \dots,x_p)^\top \in \mathbb{R}^p$. The set $E(x)$ is defined as follows:
$$E(x)=\left\{x_kx_l \mid (k,l) \in S_+ \right\}  \cup \left\{-x_kx_l \mid (k,l) \in S_- \right\}  .$$

A linear ordering $\sigma_{+}: S_{+} \rightarrow\left\{1, \ldots, p_{+}\right\}$ of $S_{+}$ is compatible with $x$ if for any two index pairs $\left(k_1, l_1\right),\left(k_2, l_2\right)$ $\in S_{+}$ such that $\sigma_{+}\left(k_1, l_1\right)<\sigma_{+}\left(k_2, l_2\right)$,  we have $x_{k_1} x_{l_1} \geqslant x_{k_2} x_{l_2}$.
A linear ordering $\sigma_{-}$ of $S_{-}$ is consistent with a linear ordering $\sigma_{+}$ of $S_{+}$ if for any two index pairs $\left(k_1, l_1\right),\left(k_2, l_2\right) \in S_{-}$ such that $\sigma_{+}\left(k_1, l_1\right)<\sigma_{+}\left(k_2, l_2\right)$, we have $\sigma_{-}\left(k_1, l_1\right)<\sigma_{-}\left(k_2, l_2\right)$.

A pair of linear orderings $\left(\sigma_{+}, \sigma_{-}\right)$ of the sets $S_{+}$ and $S_{-}$ is compatible with $x$ if $\sigma_{+}$ is compatible with $x$ and $\sigma_{-}$ is consistent with $\sigma_{+}$.
Define the finite set of pairs
$\Sigma_{\pm}(p) =\left\{(\sigma_{+}, \sigma_{-}\right) \mid \left(\sigma_{+}, \sigma_{-}\right) ~\text {compatible with }~ x=\left(x_1, \ldots, x_p\right)^{\top} ~\text{for some }  x_1\geq \dots\geq x_p>0\}.$
Note that for any $x \in \mathbb{R}^p$ with nonnegative entries, there exists at least one pair of linear orderings $\left(\sigma_{+}, \sigma_{-}\right)\in \sum_{\pm}(p)$ which is compatible with $x$.
For any pair of orderings $\left(\sigma_{+}, \sigma_{-}\right)$ and any $nm$ real numbers $\lambda_1 \geqslant \lambda_2 \geqslant \cdots \geqslant \lambda_{n m}$ assembled into a vector $\lambda$, the $p \times p$ matrix $\Lambda\left(\lambda ; \sigma_{+}, \sigma_{-}\right)$ is defined elementwise as follows:
$$\Lambda_{k l}\left(\lambda ; \sigma_{+}, \sigma_{-}\right)= \begin{cases}\lambda_{n m+1-\sigma_{+}(k, l)}, & k \leqslant l, \\ -\lambda_{\sigma_{-}(l, k)}, & k>l .\end{cases} $$

\iffalse
The matrix $\Lambda\left(\lambda ; \sigma_{+}, \sigma_{-}\right)$ has the following simple property:
Let $x \in \mathbb{R}^p$ be a vector with nonnegative entries and let $x^{\prime}$ be a permutation vector of the entries of $x$.
Suppose  $\left(\sigma_{+}, \sigma_{-}\right)$ is a pair of orderings which is compatible with $x$, and
$\left(\sigma_{+}^{\prime}, \sigma_{-}^{\prime}\right)$ is a pair of orderings which is compatible with $x^{\prime}$.
%Let further $\lambda_1 \geqslant \lambda_2 \geqslant \ldots \geqslant \lambda_{n m}$ be real numbers assembled into a vector $\lambda$,
Then $x^\top \Lambda\left(\lambda ; \sigma_{+}, \sigma_{-}\right) x =\left(x^{\prime}\right)^\top \Lambda\left(\lambda ; \sigma_{+}^{\prime}, \sigma_{-}^{\prime}\right) x^{\prime}$.
\fi

\begin{definition}\rm\cite{AbsolutelySep2007}
$\rho \in H_{+}(mn)$ is absolutely PPT if $U \rho U^{\dagger}$ is PPT for every unitary matrix $U \in U(mn)$.
\end{definition}

We consider a bipartite quantum system where the two subsystems are $m$-qudit state and $n$-qudit state, respectively. In other words, the mixed states of the composite systems are expressed as density matrices $\rho \in H_{+}(mn)$ with trace 1. Bipartite pure states can be described with vectors $|b\rangle \in \mathcal{H}_{n m}=\mathcal{H}_m \otimes \mathcal{H}_n$ of unit norm, and the density matrix of such a bipartite pure state is a rank one matrix $B=|b\rangle \langle b| \in H_{+}(mn)$. It is known that any pure state vector has a Hilbert Schmidt decomposition $|b\rangle=\sum_{k=1}^p x_k |u_k\rangle \otimes |v_k\rangle$, where
$x_1 \geqslant x_2 \geqslant \cdots \geqslant x_p \geqslant 0$ are its Schmidt coefficients,
$\left\{|u_k\rangle\right\}_{1 \leqslant k \leqslant p}$ is an orthonormal set of vectors in $\mathcal{H}_m$,
and $\left\{|v_k\rangle\right\}_{1 \leqslant k \leqslant p}$ is an orthonormal set of vectors in $\mathcal{H}_n$.
Therefore, there are local unitary transformations of $\mathcal{H}_m$ and $\mathcal{H}_n$ that bring the state vector $|b\rangle$ into its Schmidt form $\sum_{k=1}^p x_k e_k^m \otimes e_k^n$,
where $\{e_k^m\}_{1 \le k \le m}$ and $\{e_k^n\}_{1 \le k \le n}$
are the canonical basis vectors of $\mathcal{H}_m$ and $\mathcal{H}_n$, respectively.
The constraint $\langle b |b\rangle =1$ implies $\Sigma_{k=1}^p x_k^2=1$.

\begin{lemma}\rm\cite{AbsolutelySep2007}
Let $n, m$ be positive integers and let $p=\min (n, m)$. Let further $x \in \mathbb{R}^p$ be an arbitrary vector. Then there exists a vector $b \in H_{n m}$ such that the spectrum of $\left(b b^*\right)^{\Gamma}$ is given by the set $E(x)$, the remaining $p|n-m|$ eigenvalues being zero.
\end{lemma}

\begin{lemma}{\rm\cite{AbsolutelySep2007}}\label{lem1}
Suppose a mixed state $\rho$ on $\mathcal{H}_{3 n}=\mathcal{H}_3 \otimes \mathcal{H}_n$ with eigenvalues $\lambda_1, \ldots, \lambda_{3 n}$ in decreasing order.
Then $\rho$ is absolutely PPT if and only if the two linear matrix inequalities \eqref{LMI07} hold.
\begin{equation}\label{LMI07}
\begin{gathered}
\Lambda^{3}_{1}=
\begin{bmatrix}
2 \lambda_{3 n} & \lambda_{3 n-1}-\lambda_1 & \lambda_{3 n-3}-\lambda_2 \\
\lambda_{3 n-1}-\lambda_1 & 2 \lambda_{3 n-2} & \lambda_{3 n-4}-\lambda_3 \\
\lambda_{3 n-3}-\lambda_2 & \lambda_{3 n-4}-\lambda_3 & 2 \lambda_{3 n-5}
\end{bmatrix} \succeq  0, \\
\Lambda^{3}_2=
\begin{bmatrix}
2 \lambda_{3 n} & \lambda_{3 n-1}-\lambda_1 & \lambda_{3 n-2}-\lambda_2 \\
\lambda_{3 n-1}-\lambda_1 & 2 \lambda_{3 n-3} & \lambda_{3 n-4}-\lambda_3 \\
\lambda_{3 n-2}-\lambda_2 & \lambda_{3 n-4}-\lambda_3 & 2 \lambda_{3 n-5}
\end{bmatrix} \succeq   0 .
\end{gathered}
\end{equation}
\end{lemma}

\begin{lemma}{\rm\cite{AbsolutelySep2007}}\label{lem2}
Suppose a mixed state $\rho$ on $\mathcal{H}_{m n}=\mathcal{H}_m \otimes \mathcal{H}_n$ with eigenvalues $\lambda_1, \ldots, \lambda_{m n}$ in decreasing order.
Then $\rho$ is absolutely PPT if and only if for any vector $x\in \mathbb{R}^{p}$ with nonnegative and ordered entries $x_1\geq x_2\geq \cdots\geq x_p\geq 0$ there exists a pair $(\sigma_+, \sigma_-)$ of linear orderings which is compatible with $x$ such that
\begin{equation*}
  x^{\top}\Lambda(\lambda:\sigma_+, \sigma_-)x\geq 0,
\end{equation*}
that is, $\rho$ is absolutely PPT if and only if for all $(\sigma_+, \sigma_-)\in \sum_{\pm}$, we have
\begin{equation*}
  \Lambda(\lambda:\sigma_+, \sigma_-)+\Lambda(\lambda:\sigma_+, \sigma_-)^{\top}\succeq 0.
\end{equation*}
\end{lemma}

\section{Absolute separability from spectrum for ququart-qudit states}

In this section, we consider the absolute separability from spectrum for ququart-qudit states, which situation is more complex than the qutrit-qudit case. We provide the necessary and sufficient conditions for the absolute separability from spectrum for ququart-qudit states, which consists of twelve linear matrix inequalities. Given the complexity of determining these twelve linear matrix inequalities, we also present some sufficient conditions for absolute separability and non-absolute separability that rely solely on the comparison of the first few leading and last few leading eigenvalues of density matrices.

Firstly, we consider the necessary and sufficient conditions for the absolute separability of ququart-qudit states based on spectrum.

\begin{theorem}\label{thm4}
For $n \ge 4$, a mixed state $\rho$ on $\mathcal{H}_{4 n}=\mathcal{H}_4 \otimes \mathcal{H}_n$ with eigenvalues $\lambda_1, \ldots, \lambda_{4 n}$ in decreasing order.
Then $\rho$ is absolutely PPT if and only if the following twelve matrices $\Lambda_{t}, t=1,\dots,12$ are positive semidefinite:
\begin{widetext}
\begin{scriptsize}
$$\Lambda_{1}=
\begin{bmatrix}
2\lambda_{4n}& \lambda_{4n-1}-\lambda_{1}& \lambda_{4n-2}-\lambda_{2}& \lambda_{4n-3}-\lambda_{3}\\
\lambda_{4n-1}-\lambda_{1}& 2\lambda_{4n-4}& \lambda_{4n-5}-\lambda_{4}&  \lambda_{4n-6}-\lambda_{5}\\
\lambda_{4n-2}-\lambda_{2}& \lambda_{4n-5}-\lambda_{4}& 2\lambda_{4n-7}&  \lambda_{4n-8}-\lambda_{6}\\
\lambda_{4n-3}-\lambda_{3}& \lambda_{4n-6}-\lambda_{5} & \lambda_{4n-8}-\lambda_{6} & 2\lambda_{4n-9}\\
\end{bmatrix},\quad
\Lambda_{2}=
\begin{bmatrix}
2\lambda_{4n}& \lambda_{4n-1}-\lambda_{1}& \lambda_{4n-2}-\lambda_{2}& \lambda_{4n-3}-\lambda_{3}\\
\lambda_{4n-1}-\lambda_{1}& 2\lambda_{4n-4}& \lambda_{4n-5}-\lambda_{4}&  \lambda_{4n-7}-\lambda_{5}\\
\lambda_{4n-2}-\lambda_{2}& \lambda_{4n-5}-\lambda_{4}& 2\lambda_{4n-6}&  \lambda_{4n-8}-\lambda_{6}\\
\lambda_{4n-3}-\lambda_{3}& \lambda_{4n-7}-\lambda_{5} & \lambda_{4n-8}-\lambda_{6} & 2\lambda_{4n-9}\\
\end{bmatrix},
$$
$$
\Lambda_{3}=
\begin{bmatrix}
2\lambda_{4n}& \lambda_{4n-1}-\lambda_{1}& \lambda_{4n-2}-\lambda_{2}& \lambda_{4n-6}-\lambda_{4}\\
\lambda_{4n-1}-\lambda_{1}& 2\lambda_{4n-3}& \lambda_{4n-4}-\lambda_{3}&  \lambda_{4n-7}-\lambda_{5}\\
\lambda_{4n-2}-\lambda_{2}& \lambda_{4n-4}-\lambda_{3}& 2\lambda_{4n-5}&  \lambda_{4n-8}-\lambda_{6}\\
\lambda_{4n-6}-\lambda_{4}& \lambda_{4n-7}-\lambda_{5} & \lambda_{4n-8}-\lambda_{6} & 2\lambda_{4n-9}\\
\end{bmatrix},\quad
\Lambda_{4}=
\begin{bmatrix}
2\lambda_{4n}& \lambda_{4n-1}-\lambda_{1}& \lambda_{4n-3}-\lambda_{2}& \lambda_{4n-6}-\lambda_{4}\\
\lambda_{4n-1}-\lambda_{1}& 2\lambda_{4n-2}& \lambda_{4n-4}-\lambda_{3}&  \lambda_{4n-7}-\lambda_{5}\\
\lambda_{4n-3}-\lambda_{2}& \lambda_{4n-4}-\lambda_{3}& 2\lambda_{4n-5}&  \lambda_{4n-8}-\lambda_{6}\\
\lambda_{4n-6}-\lambda_{4}& \lambda_{4n-7}-\lambda_{5} & \lambda_{4n-8}-\lambda_{6} & 2\lambda_{4n-9}\\
\end{bmatrix},
$$
$$
\Lambda_{5}=
\begin{bmatrix}
2\lambda_{4n}& \lambda_{4n-1}-\lambda_{1}& \lambda_{4n-2}-\lambda_{2}& \lambda_{4n-5}-\lambda_{4}\\
\lambda_{4n-1}-\lambda_{1}& 2\lambda_{4n-3}& \lambda_{4n-4}-\lambda_{3}&  \lambda_{4n-7}-\lambda_{5}\\
\lambda_{4n-2}-\lambda_{2}& \lambda_{4n-4}-\lambda_{3}& 2\lambda_{4n-6}&  \lambda_{4n-8}-\lambda_{6}\\
\lambda_{4n-5}-\lambda_{4}& \lambda_{4n-7}-\lambda_{5} & \lambda_{4n-8}-\lambda_{6} & 2\lambda_{4n-9}\\
\end{bmatrix},\quad
\Lambda_{6}=
\begin{bmatrix}
2\lambda_{4n}& \lambda_{4n-1}-\lambda_{1}& \lambda_{4n-3}-\lambda_{2}& \lambda_{4n-5}-\lambda_{4}\\
\lambda_{4n-1}-\lambda_{1}& 2\lambda_{4n-2}& \lambda_{4n-4}-\lambda_{3}&  \lambda_{4n-7}-\lambda_{5}\\
\lambda_{4n-3}-\lambda_{2}& \lambda_{4n-4}-\lambda_{3}& 2\lambda_{4n-6}&  \lambda_{4n-8}-\lambda_{6}\\
\lambda_{4n-5}-\lambda_{4}& \lambda_{4n-7}-\lambda_{5} & \lambda_{4n-8}-\lambda_{6} & 2\lambda_{4n-9}\\
\end{bmatrix},
$$
$$
\Lambda_{7}=
\begin{bmatrix}
2\lambda_{4n}& \lambda_{4n-1}-\lambda_{1}& \lambda_{4n-2}-\lambda_{2}& \lambda_{4n-5}-\lambda_{4}\\
\lambda_{4n-1}-\lambda_{1}& 2\lambda_{4n-3}& \lambda_{4n-4}-\lambda_{3}&  \lambda_{4n-6}-\lambda_{5}\\
\lambda_{4n-2}-\lambda_{2}& \lambda_{4n-4}-\lambda_{3}& 2\lambda_{4n-7}&  \lambda_{4n-8}-\lambda_{6}\\
\lambda_{4n-5}-\lambda_{4}& \lambda_{4n-6}-\lambda_{5} & \lambda_{4n-8}-\lambda_{6} & 2\lambda_{4n-9}\\
\end{bmatrix},\quad
\Lambda_{8}=
\begin{bmatrix}
2\lambda_{4n}& \lambda_{4n-1}-\lambda_{1}& \lambda_{4n-3}-\lambda_{2}& \lambda_{4n-5}-\lambda_{4}\\
\lambda_{4n-1}-\lambda_{1}& 2\lambda_{4n-2}& \lambda_{4n-4}-\lambda_{3}&  \lambda_{4n-6}-\lambda_{5}\\
\lambda_{4n-3}-\lambda_{2}& \lambda_{4n-4}-\lambda_{3}& 2\lambda_{4n-7}&  \lambda_{4n-8}-\lambda_{6}\\
\lambda_{4n-5}-\lambda_{4}& \lambda_{4n-6}-\lambda_{5} & \lambda_{4n-8}-\lambda_{6} & 2\lambda_{4n-9}\\
\end{bmatrix},
$$
$$
\Lambda_{9}=
\begin{bmatrix}
2\lambda_{4n}& \lambda_{4n-1}-\lambda_{1}& \lambda_{4n-2}-\lambda_{2}& \lambda_{4n-4}-\lambda_{3}\\
\lambda_{4n-1}-\lambda_{1}& 2\lambda_{4n-3}& \lambda_{4n-5}-\lambda_{4}&  \lambda_{4n-7}-\lambda_{5}\\
\lambda_{4n-2}-\lambda_{2}& \lambda_{4n-5}-\lambda_{4}& 2\lambda_{4n-6}&  \lambda_{4n-8}-\lambda_{6}\\
\lambda_{4n-4}-\lambda_{3}& \lambda_{4n-7}-\lambda_{5} & \lambda_{4n-8}-\lambda_{6} & 2\lambda_{4n-9}\\
\end{bmatrix},\quad
\Lambda_{10}=
\begin{bmatrix}
2\lambda_{4n}& \lambda_{4n-1}-\lambda_{1}& \lambda_{4n-3}-\lambda_{2}& \lambda_{4n-4}-\lambda_{3}\\
\lambda_{4n-1}-\lambda_{1}& 2\lambda_{4n-2}& \lambda_{4n-5}-\lambda_{4}&  \lambda_{4n-7}-\lambda_{5}\\
\lambda_{4n-3}-\lambda_{2}& \lambda_{4n-5}-\lambda_{4}& 2\lambda_{4n-6}&  \lambda_{4n-8}-\lambda_{6}\\
\lambda_{4n-4}-\lambda_{3}& \lambda_{4n-7}-\lambda_{5} & \lambda_{4n-8}-\lambda_{6} & 2\lambda_{4n-9}\\
\end{bmatrix},
$$
$$
\Lambda_{11}=
\begin{bmatrix}
2\lambda_{4n}& \lambda_{4n-1}-\lambda_{1}& \lambda_{4n-2}-\lambda_{2}& \lambda_{4n-4}-\lambda_{3}\\
\lambda_{4n-1}-\lambda_{1}& 2\lambda_{4n-3}& \lambda_{4n-5}-\lambda_{4}&  \lambda_{4n-6}-\lambda_{5}\\
\lambda_{4n-2}-\lambda_{2}& \lambda_{4n-5}-\lambda_{4}& 2\lambda_{4n-7}&  \lambda_{4n-8}-\lambda_{6}\\
\lambda_{4n-4}-\lambda_{3}& \lambda_{4n-6}-\lambda_{5} & \lambda_{4n-8}-\lambda_{6} & 2\lambda_{4n-9}\\
\end{bmatrix},\quad
\Lambda_{12}=
\begin{bmatrix}
2\lambda_{4n}& \lambda_{4n-1}-\lambda_{1}& \lambda_{4n-3}-\lambda_{2}& \lambda_{4n-4}-\lambda_{3}\\
\lambda_{4n-1}-\lambda_{1}& 2\lambda_{4n-2}& \lambda_{4n-5}-\lambda_{4}&  \lambda_{4n-6}-\lambda_{5}\\
\lambda_{4n-3}-\lambda_{2}& \lambda_{4n-5}-\lambda_{4}& 2\lambda_{4n-7}&  \lambda_{4n-8}-\lambda_{6}\\
\lambda_{4n-4}-\lambda_{3}& \lambda_{4n-6}-\lambda_{5} & \lambda_{4n-8}-\lambda_{6} & 2\lambda_{4n-9}\\
\end{bmatrix}.
$$
\end{scriptsize}
\end{widetext}
\end{theorem}

The proof of Theorem \ref{thm4} is presented in Appendix A.
It is worth highlighting that for each $\Lambda_t$, $t=1,\ldots,12$,,
there is another $\Lambda_s$ such that $\Lambda_s$ is different from $\Lambda_t$ by swaping the two $\lambda_j$ values.
%all matrices in the set $\Lambda_t$ for $t=1,\ldots,12$ have a matrix that only swaps the positions of two elements compared to the original matrices.

In particular, it can be observed that if $\lambda_{4n-2}=\lambda_{4n-3}$, the following equalities hold:
\begin{equation*}
\Lambda_{3}=\Lambda_{4},  \Lambda_{5}=\Lambda_{6},  \Lambda_{7}=\Lambda_{8},  \Lambda_{9}=\Lambda_{10},  \Lambda_{11}=\Lambda_{12}.
\end{equation*}
Similarly, when $\lambda_{4n-6}=\lambda_{4n-7}$, the following equalities hold:
\begin{equation*}
\Lambda_{1}=\Lambda_{2}, \quad \Lambda_{5}=\Lambda_{7}, \quad \Lambda_{9}=\Lambda_{11}.
\end{equation*}
Furthermore,  if $\lambda_{4n-3}=\lambda_{4n-4}$, $\lambda_{4n-4}=\lambda_{4n-5}$ and $\lambda_{4n-5}=\lambda_{4n-6}$, then
\begin{equation*}
\Lambda_{1}=\Lambda_{9},\quad \Lambda_{5}=\Lambda_{9},\quad \Lambda_{3}=\Lambda_{5}.
\end{equation*}
In other words, if specific eigenvalues exhibit equality, the formulation of Theorem \ref{thm4} becomes more concise and easier to determine.

\begin{corollary}\label{coronly1}
Using the same notations as in Theorem \ref{thm4}.
%For $n \ge 4$, let $A$ be a self-adjoint PSD operator on $\mathcal{H}_{4 n}=\mathcal{H}_4 \otimes \mathcal{H}_n$ with eigenvalues $\lambda_1, \ldots, \lambda_{4 n}$ in decreasing order.
If the one of the following conditions holds,
\begin{enumerate}
\item[\rm (i)] $\lambda_{4n-2}=\lambda_{4n-3}$, $\lambda_{4n-6}=\lambda_{4n-7}$,
and $\Lambda_{1}$,  $\Lambda_{3}$, $\Lambda_{5}$ and $\Lambda_{9}$ are positive semidefinite matrices,
\item[\rm (ii)] $\lambda_{4n-2}=\lambda_{4n-3}$, $\lambda_{4n-5}=\lambda_{4n-6}=\lambda_{4n-7}$,
and $\Lambda_{1}$, $\Lambda_{5}$ and $\Lambda_{9}$ are positive semidefinite matrices,
\item[\rm (iii)] $\lambda_{4n-2}=\lambda_{4n-3}=\lambda_{4n-4}=\lambda_{4n-5}=\lambda_{4n-6}=\lambda_{4n-7}$,
and one of $\Lambda_t$ is a positive semidefinite matrix,
\end{enumerate}
then $A$ is absolutely PPT.
\end{corollary}

Although we have proposed necessary and sufficient conditions for the absolute separability from spectrum for ququart-qudit states, determining the twelve linear matrix inequalities $\Lambda_{t}$ is not an easy task. With this in mind, we will now focus on identifying alternative sufficient conditions that are more attainable.

We can derive a class of sufficient conditions that are easy to determine and implement, which only rely on the first few leading and last few leading eigenvalues of the density matrix.

\begin{theorem}\label{thm4PPT}
Using the same notations as in Theorem \ref{thm4}.
%For $n \ge 4$, let $A$ be a self-adjoint PSD operator on $\mathcal{H}_{4 n}=\mathcal{H}_4 \otimes \mathcal{H}_n$ with eigenvalues $\lambda_1, \ldots, \lambda_{4 n}$ in decreasing order.
If the following inequality holds:
\begin{equation}\label{thm6}
  \lambda_{4n}+\lambda_{4n-1}+\lambda_{4n-2}+\lambda_{4n-3}\geq \lambda_1+\lambda_2+\lambda_3,
\end{equation}
%where $\lambda_1, \ldots, \lambda_{4n}$ are the eigenvalues of $A$ in decreasing order,
then $\rho$ is absolutely PPT. %See Appendix C for the proof.
\end{theorem}
{\bf Proof:} It is worth noting that if the inequality \eqref{thm6} holds, we have the following chain of inequalities:
\begin{eqnarray*}
% \nonumber % Remove numbering (before each equation)
   && 2\lambda_{4n}+\lambda_{4n-1}+\lambda_{4n-2}+\lambda_{4n-3} \\
  &\geq  & \lambda_{4n}+\lambda_{4n-1}+\lambda_{4n-2}+\lambda_{4n-3} \\
  &\geq  & \lambda_1+\lambda_2+\lambda_3 ,
\end{eqnarray*}
which yields
\begin{equation}\label{thmeq02}
  2\lambda_{4n}\geq \lambda_1-\lambda_{4n-1}+\lambda_2-\lambda_{4n-2}+\lambda_3-\lambda_{4n-3}.
\end{equation}

It is noting that  $\Lambda_{t}$, $t=1,\dots,12$, have the largest diagonal elements in absolute value and the smallest sum of the absolute values of the off-diagonal elements
in their first row. Based on equality \eqref{thmeq02},
then all matrices $\Lambda_{t}$, mentioned in Theorem \ref{thm4} are diagonal dominant and positive semidefinite. Therefore, $A \in \mathcal{H}_{4n}$ is absolutely PPT. This completes the proof. $\Box$

Given the inequality \eqref{thm6} and the fact that $\lambda_{4n-1}+\lambda_{4n-2}\leq \lambda_2+\lambda_3$, we can immediately derive the following result.

\begin{corollary}
Using the same notations as in Theorem \ref{thm4}.
%For $n \ge 4$, let $A$ be a self-adjoint PSD operator on $\mathcal{H}_{4 n}=\mathcal{H}_4 \otimes \mathcal{H}_n$ with eigenvalues $\lambda_1, \ldots, \lambda_{4 n}$ in decreasing order.
%Consider a self-adjoint positive semidefinite (PSD) operator $A$ on the Hilbert space $\mathcal{H}_{4n}=\mathcal{H}_4 \otimes \mathcal{H}_n$ with $n \geq 4$. Let $\lambda_1, \ldots, \lambda_{4n}$ denote the eigenvalues of $A$ in decreasing order.
If the inequality
\begin{equation*}
\lambda_{4n}+\lambda_{4n-1}\geq \lambda_1
\end{equation*}
holds, then $\rho$ is absolutely PPT.
\end{corollary}

%\begin{remark}{\bf TBC}
%It is worth noting that if the inequality \eqref{thm6} holds, we have the following chain of inequalities:
%\begin{eqnarray*}
%% \nonumber % Remove numbering (before each equation)
%   && 2\lambda_{4n}+\lambda_{4n-1}+\lambda_{4n-2}+\lambda_{4n-3} \\
%  \geq  && \lambda_{4n}+\lambda_{4n-1}+\lambda_{4n-2}+\lambda_{4n-3} \\
%  \geq  && \lambda_1+\lambda_2+\lambda_3.
%\end{eqnarray*}
%%$
%%2\lambda_{4n}+\lambda_{4n-1}+\lambda_{4n-2}+\lambda_{4n-3}\geq \lambda_{4n}+\lambda_{4n-1}+\lambda_{4n-2}+\lambda_{4n-3}\geq \lambda_1+\lambda_2+\lambda_3.
%%$
%Then all matrices $\Lambda_{t}, t=1,\dots,12$, mentioned in Theorem \ref{thm4} are diagonal dominant and positive semidefinite. Therefore, $A \in \mathcal{H}_{4n}$ is absolutely PPT.
%\end{remark}

On the other hand, let us consider the conditions under which $\rho \in \mathcal{H}_{4n}$ is not absolutely PPT. It is known that even if only some of the rows of a matrix are diagonally dominant, it may still be positive semidefinite. However, if none of the rows are diagonally dominant, the matrix is not positive semidefinite.  Here we have the following result.

\iffalse
For example,
we assuming that $A=(a_{ij})\in M_q(\mathbb{C})$ is positive semidefinite, if none of the rows of a symmetric matrix $A$ are diagonally dominant, we have:
$$a_{ii}\leq \sum\limits_{i\neq j}|a_{ij}|=-\sum\limits_{i\neq j}a_{ij}, i=1,\cdots,q,$$
This implies that $\sum\limits_{i= 1}^{q}a_{ij}\leq0$. Let $x=(1,\cdots,1)^{\top}\in \mathbb{R}^{q}$ be a vector with all elements equal to 1. Then we have:
$$x^{\top}Ax=\sum\limits_{i=1}^{q}\sum\limits_{j=1}^{q}a_{ij}x_{i}x_{j}=\sum\limits_{i=1}^{q}\sum\limits_{j=1}^{q}a_{ij}\leq 0,$$
which contradicts the positive semidefiniteness of $A$.
%Before delving into that, let us introduce the following lemma.

\begin{lemma}\label{lemmaND}
For a symmetric square matrix $A$ of dimension $q$, with positive diagonal elements and negative off-diagonal elements, if none of the rows of the matrix are diagonally dominant, then $A$ is not positive semidefinite.
\end{lemma}
\fi

\begin{theorem}\label{thm4NPPT}
Using the same notations as in Theorem \ref{thm4}.
%For $n \ge 4$, let $A$ be a self-adjoint PSD operator on $\mathcal{H}_{4 n}=\mathcal{H}_4 \otimes \mathcal{H}_n$ with eigenvalues $\lambda_1, \ldots, \lambda_{4 n}$ in decreasing order.
%Consider a self-adjoint positive semidefinite (PSD) operator $A$ on the Hilbert space $\mathcal{H}_{4n}=\mathcal{H}_4 \otimes \mathcal{H}_n$ with $n \geq 4$. Let $\lambda_1, \ldots, \lambda_{4 n}$ be the eigenvalues of $A$ in decreasing order.
If the following inequality holds:
\begin{equation}\label{eq3.06}
  \lambda_3+\lambda_5+\lambda_6 \geq 2\lambda_{4n-9}+\lambda_{4n-8}+\lambda_{4n-6}+\lambda_{4n-3},
\end{equation}
then $A$ is not absolutely PPT.
\end{theorem}
{\bf{Proof}}:
If \eqref{eq3.06} holds, then
\begin{equation*}
  2\lambda_{4n-9}\leq (\lambda_3-\lambda_{4n-3})+(\lambda_5-\lambda_{4n-6})+(\lambda_6-\lambda_{4n-8}),
\end{equation*}
which shows that the fourth row of the matrix $\Lambda_{1} $ is not diagonally dominant.

It is noting that  $\Lambda_{1}$ has the largest diagonal elements in absolute value and the smallest sum of the absolute values of the off-diagonal elements
in its fourth row. Therefore, none of the rows of matrix $\Lambda_{1} $ are diagonally dominant. We can conclude that $\rho$ is not absolutely PPT.
This completes the proof. $\Box$

\iffalse
%Considering the eigenvalues $\lambda_1, \ldots, \lambda_{4 n}$ in decreasing order and the construction of $\Lambda_{t} (t=1,\cdots,12)$,
It is worth noting that we can construct some boundary matrices such that if the boundary matrix does not have diagonally dominant rows, the remaining matrices will not have diagonally dominant rows. For instance, in Theorem \ref{thm4}, if the last rows of matrices $\Lambda_{3}$ and $\Lambda_{4}$ fail to satisfy diagonal dominance, that is,
\begin{equation*}
  \lambda_4+\lambda_5+\lambda_6 \geq 2\lambda_{4n-9}+\lambda_{4n-8}+\lambda_{4n-7}+\lambda_{4n-6},
\end{equation*}
which implies that none of the matrices $\Lambda_{t}$ have diagonally dominant rows.
\fi

\iffalse
It is not difficult to observe that the following inequalities hold:
$\lambda_1+\lambda_2+\lambda_3\geq \lambda_4+\lambda_5+\lambda_6,  2\lambda_{4n-9}+\lambda_{4n-8}+\lambda_{4n-7}+\lambda_{4n-6}\geq \lambda_{4n}+\lambda_{4n-1}+\lambda_{4n-2}+\lambda_{4n-3}.$
These inequalities imply that Theorem \ref{thm4PPT} and Theorem \ref{thm4NPPT} do not overlap, which is advantageous for partitioning the sets of absolutely separable and non-absolutely separable states.
\fi

\section{Absolute separability from spectrum for qudit-qudit states}

In this section, we investigate the absolute separability of bipartite states in the general Hilbert space $\mathcal{H}_{mn}=\mathcal{H}_m \otimes \mathcal{H}_{n}$, where $n\geq m$.
We aim to extend some of the findings for $\mathcal{H}_{3n}$ and $\mathcal{H}_{4n}$ to the general case $\mathcal{H}_{mn}$, where $n\geq m\geq 4$.
In this context, we seek sufficient conditions for absolute separability and non-absolute separability in the Hilbert space $\mathcal{H}_{mn}$, which depend solely on $2m-1$ eigenvalues of the density matrices. Additionally, we can derive distance bounds for the eigenvalues and purity bounds for general absolutely separable states.

First, we show that Theorem \ref{thm4PPT} can be extended to the general case of the Hilbert space $\mathcal{H}_{mn}$ for $n\geq m$. This extension significantly enhances our understanding of absolutely separable states in bipartite state spaces.

\begin{theorem}\label{neccery_mn}
For $n  \ge m \ge 2$ and given a mixed state $\rho$ on $\mathcal{H}_{m n}=\mathcal{H}_m \otimes \mathcal{H}_n$ with eigenvalues $\lambda_1, \ldots, \lambda_{m n}$ in decreasing order.
%Consider a self-adjoint positive semidefinite (PSD) operator $A$ on the Hilbert space $\mathcal{H}_{mn}$ with $n \geq m\geq 2$. Let $\lambda_1, \ldots, \lambda_{m n}$ be the eigenvalues of $A$ in decreasing order.
If the following inequality holds:
\begin{equation}\label{thmmn}
  \lambda_{mn}+\lambda_{mn-1}+\cdots+\lambda_{m(n-1)+1}\geq \lambda_1+\lambda_2+\cdots+\lambda_{m-1},
\end{equation}
then $\rho$ is absolutely PPT.
\end{theorem}
{\bf Proof:} Let $n\geq m\geq 2$. Then $p=m$, $p_{+}=\frac{m(m+1)}{2}$, and $p_{-}=\frac{m(m-1)}{2}$. We need to determine the set $\sum_{\pm}(m)$. Without loss of generality, we assume that $\alpha_{1}\geq \alpha_{2} \geq \cdots  \geq \alpha_{m}\geq 0$.
There are always such orderings for the numbers $\{\alpha_{i} \alpha_{j}\}_{1\le i,j\le m}$:
\begin{small}
\begin{equation}\label{linearoder1}
  \alpha_{1}^{2}\geq \alpha_{1} \alpha_{2}\geq \cdots\geq  \alpha_{1}\alpha_{m} \geq  \alpha_{2}^{2}\geq \alpha_{2}\alpha_{3}\geq  \cdots\geq \alpha_{m}^{2}.
\end{equation}
\end{small}
%$$\alpha_{1}^{2}\geq \alpha_{1} \alpha_{2}\geq \cdots\geq  \alpha_{1}\alpha_{m} \geq  \alpha_{2}^{2}\geq \alpha_{2}\alpha_{3}\geq  \cdots\geq \alpha_{m}^{2}.$$

We construct the corresponding matrix $\Lambda^{m}_{1}$ as follows:
\iffalse
\begin{equation*}
\begin{bmatrix}
\lambda_{mn}& \lambda_{mn-1}& \lambda_{mn-2}&  \cdots &\lambda_{m(n-1)+1}\\
-\lambda_{1}& \cdots & \cdots &  \cdots & \cdots\\
-\lambda_{2}& \cdots& \cdots&   \cdots & \cdots\\
\vdots & \vdots & \vdots &  \vdots & \vdots\\
-\lambda_{m-1} & \cdots&  \cdots& \cdots  & \lambda_{mn-(\frac{m(m+1)}{2}-1)}
\end{bmatrix},
\end{equation*}
\fi
\begin{small}
\begin{equation*}
\begin{bmatrix}
2\lambda_{mn}& \lambda_{mn-1}-\lambda_{1}&   \cdots &\lambda_{m(n-1)+1}-\lambda_{m-1}\\
\lambda_{mn-1}-\lambda_{1}& \ddots &   \cdots & \cdots\\
\vdots & \vdots & \ddots &   \vdots\\
\lambda_{m(n-1)+1}-\lambda_{m-1} & \cdots&   \cdots  & 2\lambda_{mn-(\frac{m(m+1)}{2}-1)}
\end{bmatrix}.
\end{equation*}
\end{small}

\iffalse
We could construct a symmetric matrix $\Lambda^{m}_{1}$ with the following first row and column:
$\Lambda^{m}_{1}(1,1)=\lambda_{mn}, \Lambda^{m}_{1}(1,2)=\Lambda^{m}_{1}(2,1)=\lambda_{mn-1}-\lambda_1, \cdots, \Lambda^{m}_{1}(1,m)=\Lambda^{m}_{1}(m,1)=\lambda_{m(n-1)+1}-\lambda_{m-1}.$

Consequently, we have
\begin{eqnarray*}
% \nonumber % Remove numbering (before each equation)
   &&  R_{1}(\Lambda^{m}_{1}) \\
  = &&  [\lambda_{mn}x_1+(\lambda_{mn-1}-\lambda_1)x_2\\
   && +\cdots+(\lambda_{m(n-1)+1}-\lambda_{m-1})x_m]x_1\\
   \geq && 0.
\end{eqnarray*}

If the following inequality holds %$$\lambda_{mn}+\lambda_{mn-1}+\cdots+\lambda_{m(n-1)+1}\geq \lambda_1+\lambda_2+\cdots+\lambda_{m-1},$$
\begin{eqnarray*}
&&\lambda_{mn}+\lambda_{mn-1}+\cdots+\lambda_{m(n-1)+1}\\
\geq &&\lambda_1+\lambda_2+\cdots+\lambda_{m-1},
\end{eqnarray*}
then we have
\begin{eqnarray*}
&&\lambda_{mn}\frac{x_1}{x_m}+\lambda_{mn-1}+\cdots+\lambda_{m(n-1)+1}\\
\geq &&\lambda_1+\lambda_2+\cdots+\lambda_{m-1},
\end{eqnarray*}
which yields
\begin{eqnarray*}
&&\lambda_{mn}x_1+(\lambda_{mn-1}-\lambda_1)x_m\\
&&~~+\cdots+(\lambda_{m(n-1)+1}-\lambda_{m-1})x_m\\
\geq && 0.
\end{eqnarray*}
This shows that
\begin{eqnarray*}
% \nonumber % Remove numbering (before each equation)
   &&  [\lambda_{mn}x_1+(\lambda_{mn-1}-\lambda_1)x_2\\
   && ~~+ \cdots+(\lambda_{m(n-1)+1}-\lambda_{m-1})x_m]x_1\\
  \geq  && R_{1}(\Lambda^{m}_{1})\geq 0.
\end{eqnarray*}
\fi

Here the set of matrices $\Omega$ is corresponding to the set $\sum_{\pm}(m)$ such that every matrix $\Lambda^{m}_{k} (k=1,2,\dots)$ in $\Omega$ is corresponding to one pair of linear orderings in $\sum_{\pm}(m)$. In particular, the matrix $\Lambda^{m}_{1}$ is corresponding to linear ordering \eqref{linearoder1}.

It is not difficult to find that the first row and column of matrix $\Lambda^{m}_{1}$ exhibit the smallest sum of the absolute values diagonal and the largest the absolute values of the off-diagonal element.
%If the first row of matrix $\Lambda^{m}_{1}$ satisfies the diagonally dominant condition,  then all matrices $\Lambda^{m}_{k}(k=1,2,\dots )$ are diagonally dominant matrices.
%where the set of matrices $\Lambda^{m}$ is corresponding to the set $\sum_{\pm}(m)$ such that every matrix $\Lambda^{m}_{k} (k=1,2,\dots)$ in $\Lambda^{m}$ is corresponding to one pair of linear orderings in $\sum_{\pm}(m)$.
If the inequality \eqref{thmmn} holds, then the first row of matrix $\Lambda^{m}_{1}$ satisfies the diagonally dominant condition. Furthermore, all matrices $\Lambda^{m}_{k}$, $k=1,2,\dots$, are diagonally dominant matrices, and they are positive semidefiniteness.
Therefore, $\rho$ is absolutely positive partial transpose (PPT). The proof is completed. $\Box$
%These conditions establish the validity of inequality \eqref{thmmn}, thereby confirming that $\rho$ is absolutely positive partial transpose (PPT). The proof is completed. $\Box$

%{\color{blue}
%Given that the eigenvalues ($\lambda_1, \ldots, \lambda_{mn}$) of $A$ are arranged in decreasing order, we have the inequality:
%$$\lambda_{mn-2}+\cdots+\lambda_{m(n-1)+1}\leq \lambda_2+\cdots+\lambda_{m-1}.$$}
%See Appendix B for the proof.
Based on the above inequality \eqref{thmmn}, we can derive the following conclusion immediately.

\begin{corollary}\label{pro23}
Using the same notations as in Theorem \ref{neccery_mn}.
%Consider a self-adjoint positive semidefinite (PSD) operator $A$ on the Hilbert space $\mathcal{H}_{mn}$ with $n\geq m \geq 2$.
%Let $\lambda_1, \ldots, \lambda_{m n}$ be the eigenvalues of $A$ in decreasing order.
If the following inequality holds:
\begin{equation}\label{mn21}
\lambda_{mn}+\lambda_{mn-1}\geq \lambda_1.
\end{equation}
then $\rho$ is absolutely PPT.
\end{corollary}

\begin{remark}
It is worth noting that Corollary \ref{pro23} represents a significant improvement over \eqref{mneq1}, which is a result found in the literature \cite{JIVULESCU2015276,Filippov2017}.
In other words, the existing conclusion \eqref{mneq1} can be considered as a special case of our more general conclusion (Theorem \ref{neccery_mn} and Corollary \ref{pro23}).
\end{remark}

\begin{remark}
When $m=2$, we can easily observe the following inequalities:
\begin{equation*}
\lambda_1\leq \lambda_{2n}+\lambda_{2n-1}\leq \lambda_{2n-1}+2\sqrt{\lambda_{2n-2}\lambda_{2n}},
\end{equation*}
which demonstrate that the bounds \eqref{mn21} are tighter than the bound \eqref{ieqAS13}.

In the case  $m=3$, we have $\lambda_{3n}+\lambda_{3n-1}\geq \lambda_1$, which implies $\lambda_{3n}+\lambda_{3n-2}\geq \lambda_2$. Consequently, we obtain the following inequality:
\begin{equation*}
2\lambda_{3n}+\lambda_{3n-1}+\lambda_{3n-2}\geq \lambda_{3n}+\lambda_{3n-1}+\lambda_{3n-2}\geq \lambda_1+\lambda_2.
\end{equation*}
Since the eigenvalues $\lambda_1, \ldots, \lambda_{3n}$ of $\rho$ are arranged in decreasing order, then both $\Lambda^{3}_{1}$ and $\Lambda^{3}_{2}$ in Lemma \ref{lem1} are diagonally dominant matrices, which indicate that $\Lambda^{3}_{1}$ and $\Lambda^{3}_{2}$ are positive semidefinite. In other words, if the inequality \eqref{mn21} holds, the linear matrix inequalities in Lemma \ref{lem1} are satisfied. Therefore, $\rho$ must be PPT.
\end{remark}

Corollary \ref{pro23} serves as a more precise sufficient condition compared to the existing conclusions in lower dimensions ($m=2,3$). Furthermore, it offers straightforward and practical criteria for identifying general absolutely separable states in the Hilbert space $\mathcal{H}_{mn}$ with $m\leq n$. These criteria also circumvent the NP-hard problem of determining the positive semidefiniteness of large-scale parameter matrices in high dimensions.

%Given that the eigenvalues of $A$ are arranged in decreasing order,
In addition, the inequality \eqref{thmmn} provides lower bounds on the eigenvalues of density matrix $\rho$ for being an absolutely separable state.

\begin{proposition}
Using the same notations as in Theorem \ref{neccery_mn}.
If the inequality \eqref{thmmn} holds, indicating that $\rho$ is absolutely PPT, then we have
\begin{equation*}
\frac{\lambda_{m(n-1)+1}}{\lambda_{m-1}}\geq \frac{m-1}{m}.
\end{equation*}
\end{proposition}
{\bf Proof:} Since the eigenvalues $\lambda_1, \ldots, \lambda_{mn}$ of $\rho$ are arranged in decreasing order, obviously there are $\lambda_{m(n-1)+1}\geq \cdots \geq \lambda_{mn}$ and $\lambda_{1}\geq \cdots \geq \lambda_{m-1}$. Based on inequality \eqref{thmmn}, the desired conclusion immediately holds. This completes the proof. $\Box$

Similarly, the inequality \eqref{thmmn} also provides some lower bounds on the purity of absolutely separable states.

\begin{proposition}\label{sec4pro3}
Using the same notations as in Theorem \ref{neccery_mn}.
If the inequality \eqref{thmmn} holds, indicating that $\rho$ is absolutely PPT, there are as follows:
\begin{equation}\label{sectreq1}
\left(\frac{m-1}{m} \right)^2\lambda_{m-1}^2\leq \frac{\operatorname{tr}(\rho^2)-\lambda_1^2}{mn-1},
\end{equation}
and
\begin{equation}\label{sectreq2}
\operatorname{tr}(\rho^2)\geq \left( \left(\frac{m-1}{m} \right)^2(mn-1)+1\right)\lambda_{m-1}^2.
\end{equation}
\end{proposition}
{\bf Proof:}
Based on the Cauchy-Schwarz inequality, we can establish the following relationships:
\begin{eqnarray*}
% \nonumber % Remove numbering (before each equation)
   && (\lambda_{mn}+\lambda_{mn-1}+\cdots+\lambda_{m(n-1)+1})^2 \\
  &\leq  & m(\lambda_{mn}^2+\lambda_{mn-1}^2+\cdots+\lambda_{m(n-1)+1}^2) \\
  &\leq & m\frac{m}{nm-1}\sum\limits_{i=2}^{mn}\lambda_{i}^{2}
  = m^{2} \frac{\operatorname{tr}(\rho^2)-\lambda_1^2}{mn-1}.
\end{eqnarray*}
Moreover, the inequality \eqref{thmmn} implies that
\begin{eqnarray*}
% \nonumber % Remove numbering (before each equation)
   && (\lambda_{mn}+\lambda_{mn-1}+\cdots+\lambda_{m(n-1)+1})^2 \\
  &\geq  & (\lambda_1+\lambda_2+\cdots+\lambda_{m-1})^2 \\
  &\geq & (m-1)^2\lambda_{m-1}^2.
\end{eqnarray*}
Combining the above two inequalities, we obtain the inequality \eqref{sectreq1}.
According to $\lambda_1\geq \frac{1}{mn}$, then
\begin{eqnarray*}
\operatorname{tr}(\rho^2)
&\geq & \left(\frac{m-1}{m} \right)^2(mn-1 )\lambda_{m-1}^2+\left(\frac{1}{mn} \right)^2\\
&\geq &\left( \left(\frac{m-1}{m} \right)^2(mn-1)+1 \right)\lambda_{m-1}^2,
\end{eqnarray*}
which implies the inequality \eqref{sectreq2}. Thus, the proof is completed. $\Box$

%There exists a sufficient condition for absolute separability for general bipartition states in Hilbert space $\mathcal{H}_{mn}=\mathcal{H}_m \otimes \mathcal{H}_n$, which is derived from the observation that states $\rho$ with a sufficiently low purity $\operatorname{tr}(\rho^2)$ are separable \cite{Ball02, Ball98,Ball05,AbsolutelySep2001}. If the state $\rho$ in Hilbert space $\mathcal{H}_{mn}$ satisfies the following
%\begin{equation}\label{eqball02}
%   \operatorname{tr}(\rho^2)=\sum_{k=1}^{mn}\lambda_{k}^{2}\leq \frac{1}{mn-1},
%\end{equation}
%then $\rho$ is separable. Due to the invariance of the Frobenius norm under unitary rotations $U\rho U^{*}$, all states within the separable ball \eqref{eqball02} are guaranteed to be absolutely separable, that is, $\rho$ is absolutely separable.

It is known that a state $\rho$ with a sufficiently low purity $\operatorname{tr}(\rho^2)$ is separable \cite{Ball02, Ball98, Ball05, AbsolutelySep2001,Hildebrand072} in the Hilbert space $\mathcal{H}_{mn}$.
%In the Hilbert space $\mathcal{H}_{mn}=\mathcal{H}_m \otimes \mathcal{H}_n$, there exists a sufficient condition for absolute separability of general bipartition states.
%This condition is derived from the observation that states $\rho$ with a sufficiently low purity $\operatorname{tr}(\rho^2)$ are separable \cite{Ball02, Ball98, Ball05, AbsolutelySep2001,Hildebrand072}.
If a state $\rho$ satisfies the following inequality \cite{Filippov2017}:
\begin{equation}\label{eqball02}
\operatorname{tr}(\rho^2)=\sum_{k=1}^{mn}\lambda_{k}^{2}\leq \frac{1}{mn-1},
\end{equation}
then it is separable. Moreover, due to the invariance of the Frobenius norm under unitary rotations $U\rho U^{\dag}$, all states within the separable ball defined by Equation \eqref{eqball02} are guaranteed to be absolutely separable. In other words, the state $\rho$ is absolutely separable.

Inequalities for eigenvalues, such as \eqref{mneq1} and \eqref{mn21},  impose a restriction on the purity of absolutely separable states, indicating a limitation on their purity. Building upon this, the literature \cite{Filippov2017} has derived the following conclusion based on the condition \eqref{mneq1}.

\begin{proposition}{\rm\cite{Filippov2017}}\label{purity2017}
Let $\rho$ be a mixed state on the Hilbert space $\mathcal{H}_{mn}$ with $m\leq n$.
%Denote the eigenvalues of $A$ in decreasing order as $\lambda_1, \ldots, \lambda_{mn}$.
If $\rho$ is absolutely positive partial transpose (PPT), the following inequalities hold:
%\begin{scriptsize}
\begin{equation*}
  1+\sqrt{\frac{k \operatorname{tr}(\rho^2)-1}{k-1}} \leqslant 3 k \sqrt{\frac{\operatorname{tr}(\rho^2)}{mn+8}} , \ \hbox{if}\ \frac{1}{k} \leqslant \operatorname{tr}(\rho^2) \leqslant \frac{1}{k-1},
\end{equation*}
%\end{scriptsize}
for some $k=2, \ldots, mn,$ and
\begin{equation*}
  \operatorname{tr}(\rho^2)\leq \frac{9}{mn+8}.
\end{equation*}
%Here $\frac{1}{k} \leqslant \operatorname{tr}(\rho^2) \leqslant \frac{1}{k-1}, k=2,3, \ldots$
\end{proposition}

In a similar manner, we can provide a quantitative characterization of the maximum ball that encompasses the set of all absolutely separable states. This characterization is based on the conditions stated in \eqref{mn21}.

\begin{proposition}\label{purity}
Let $\rho$ be a mixed state on the Hilbert space $\mathcal{H}_{mn}$ with $m\leq n$.
%Denote the eigenvalues of $A$ in decreasing order as $\lambda_1, \ldots, \lambda_{mn}$.
If $\rho$ is absolutely positive partial transpose (PPT), the following inequalities hold:
\begin{small}
\begin{equation}\label{purityeq1}
  1+\sqrt{\frac{k \operatorname{tr}(\rho^2)-1}{k-1}} \leqslant 2 k \sqrt{\frac{\operatorname{tr}(\rho^2)}{mn+3}} , \ \hbox{if}\ \frac{1}{k} \leqslant \operatorname{tr}(\rho^2) \leqslant \frac{1}{k-1},
\end{equation}
\end{small}
for some $k=2, \ldots, mn,$ and
\begin{equation}\label{purityeq2}
  \operatorname{tr}(\rho^2)\leq \frac{4}{mn+3}.
\end{equation}
\end{proposition}
{\bf Proof:}
It is noting that
$$(\lambda_{mn}+\lambda_{mn-1})^2\leq 2(\lambda_{mn}^2+\lambda_{mn-1}^2) \leq 4\frac{\operatorname{tr}(\rho^2)-\lambda_1^2}{mn-1}.$$
Consequently, if there is
\begin{equation}\label{Apfeq1}
  \lambda_{1}^2>4\frac{\operatorname{tr}(\rho^2)-\lambda_1^2}{mn-1},
\end{equation}
then $\lambda_{1}> \lambda_{mn}+\lambda_{mn-1}$, which is contradicted with  \eqref{mn21}.

If the purity $\operatorname{tr}(\rho^2)$ of a system is known, it is possible to establish a lower bound for the maximal eigenvalue $\lambda_{1}$. This lower bound can be determined using the method of Lagrange multipliers with constraints $\sum\limits_{i=1}^{mn}\lambda_i=1$ and $\lambda_{1}\geq \lambda_{2}\geq \cdots \geq \lambda_{mn}\geq 0$.
Additionally, the eigenvalue $\lambda_{1}$ is minimal when both $\lambda_{1}=\cdots =\lambda_{k-1}$ and $\lambda_{k+1}=\lambda_{k+2}=\cdots =\lambda_{mn} = 0$ are satisfied for some $1<k\leq mn$.
That is to say, there are $\lambda_{k}=1-(k-1)\lambda_{1}$ and $\operatorname{tr}(\rho^2)=(k-1)\lambda_{1}^{2}+[1-(k-1)\lambda_{1}]^{2}$.
If $\frac{1}{k}\leq \operatorname{tr}(\rho^2)\leq \frac{1}{k-1}$, then the minimal largest eigenvalue can be expressed as follows:
\begin{equation}\label{Apfeq2}
  \min\lambda_{1}=\frac{1}{k}\bigg(1+\sqrt{\frac{k\operatorname{tr}(\rho^2)-1}{k-1}}\bigg).
\end{equation}
By substituting $\min\lambda_{1}$ into equation \eqref{Apfeq1}, we derive a converse to inequality \eqref{purityeq1}. Therefore, inequality \eqref{purityeq1} is an necessary condition for absolute separability.
Formula \eqref{purityeq2} can be derived from inequality \eqref{purityeq1} and represents a hyperbola that intersects all the critical points of $\operatorname{tr}(\rho^2)$ as a function of $mn$:
$$\lambda_{1}^2\leq 4\frac{\operatorname{tr}(\rho^2)-\lambda_1^2}{mn-1}\quad \Longrightarrow\quad \operatorname{tr}(\rho^2)\leq \frac{4}{mn+3}.$$
The proof is completed. $\Box$

Proposition \ref{purity} indicates that if the purity of the density operator, $\operatorname{tr}(\rho^2)$, is less than or equal to $\frac{4}{mn+3}$, then a state $\rho\in \mathcal{H}_{mn}$ is  absolutely separable with respect to any partition $m|n$ (where $mn\geq 4$ and $m,n\geq 2$).
%\footnote{See Appendix C for the proof of Proposition \ref{purity}.}

\begin{remark}
It is evident that the following inequalities hold:
\begin{equation*}
2k \sqrt{\frac{\operatorname{tr}(\rho^2)}{mn+3}} \leq 3k \sqrt{\frac{\operatorname{tr}(\rho^2)}{mn+8}} \quad\hbox{and}\quad \frac{4}{mn+3} \leq \frac{9}{mn+8}.
\end{equation*}
These inequalities demonstrate how our results, as presented in Proposition \ref{purity}, enhance the findings of Proposition \ref{purity2017} ( \cite[Proposition 1] {Filippov2017}).
\end{remark}

%In short, our upper bounds on the condition of absolute separability using purity will be more precise than that in Reference in \cite{Filippov2017}.

In a similar manner to the case of $\mathcal{H}_{4n}$ presented in Section III, we now consider the conditions under which $A\in \mathcal{H}_{mn}$ is not absolutely PPT.

%To achieve this, we aim to identify certain boundary matrices among the set of all matrices $\Lambda^{m}_{k}$ (where $k$ denotes the number of matrices). These boundary matrices will serve as a representation of our concept. In other words, if a boundary matrix satisfies a given condition, then all matrices $\Lambda^{m}_{k}$ will also satisfy that condition.

To accomplish this, we aim to identify certain matrices from the set of matrices $\Omega$ that every matrix $\Lambda^{m}_{k}$, $k=1,2,\dots$, in this set is corresponding to one pair of linear orderings in the set $\sum_{\pm}(m)$.
The objective is to find a matrix $\Lambda^{m}_{k}$ that, if one of the rows satisfies the desired condition of non-diagonal dominance, guarantee that all rows of matrix $\Lambda^{m}_{k}$ will also satisfy it. This approach allows us to establish the conditions for the non-absolute PPT across the entire set of matrices $\Omega$.

\begin{theorem}\label{thmMNPPT}
Let $\rho$ be a mixed state on the Hilbert space $\mathcal{H}_{mn}$ with $m\leq n$.
%Denote the eigenvalues of $A$ in decreasing order as $\lambda_1, \ldots, \lambda_{mn}$.
If the following condition holds:
\begin{eqnarray}\label{MNNPPT}
   && \lambda_{\frac{(m-1)(m-2)}{2}+1}+\lambda_{\frac{(m-1)(m-2)}{2}+2}+\cdots+\lambda_{\frac{(m-1)m}{2}} \nonumber\\
  &\geq & \lambda_{mn-\frac{m(m-1)}{2}}+\lambda_{mn-\frac{m(m-1)}{2}+1}+\cdots \\
   &&  +\, \lambda_{mn-(\frac{m(m+1)}{2}-2)}+2\lambda_{mn-(\frac{m(m+1)}{2}-1)} \nonumber,
\end{eqnarray}
%\begin{widetext}
%\begin{tiny}
%\begin{equation}\label{MNNPPT}
%\lambda_{\frac{(m-1)(m-2)}{2}+1}+\lambda_{\frac{(m-1)(m-2)}{2}+2}+\cdots+\lambda_{\frac{(m-1)m}{2}} \geq \lambda_{mn-\frac{m(m-1)}{2}}+\lambda_{mn-\frac{m(m-1)}{2}+1}+\cdots +\lambda_{mn-(\frac{m(m+1)}{2}-2)}+2\lambda_{mn-(\frac{m(m+1)}{2}-1)},
%\end{equation}
%\end{tiny}
%\end{widetext}
then $\rho$ is not absolutely PPT.
\end{theorem}

Theorem \ref{thmMNPPT} indicates that if the eigenvalues of the density matrix $\rho$ satisfy inequality \eqref{MNNPPT}, they are not absolutely separable.
The proof of Theorem \ref{thmMNPPT} is presented in Appendix B.
%\footnote{\footnotesize The proof of Theorem \ref{thmMNPPT} is presented in Appendix B.}

\begin{remark}
For the case when $m=2$, the inequality \eqref{MNNPPT} can be expressed as:
\begin{equation}\label{2NNPPT28}
\lambda_{1}\geq \lambda_{2n}+\lambda_{2n-1}+2\lambda_{2n-2},
\end{equation}
it follows that
\begin{eqnarray*}%\label{2NNPPT28}
\lambda_{1} &\geq & \lambda_{2n}+\lambda_{2n-1}+2\lambda_{2n-2}  \\
&\geq & \lambda_{2n-1}+2\sqrt{\lambda_{2n-2}\lambda_{2n}}+\lambda_{2n-2}\\
&\geq & \lambda_{2n-1}+2\sqrt{\lambda_{2n-2}\lambda_{2n}},
\end{eqnarray*}
which violates the necessary and sufficient condition \eqref{mneq1} of absolute separability.
This implies that if \eqref{2NNPPT28} holds, the mixed state $\rho$ in the Hilbert space $\mathcal{H}_{2n}$, where $n\geq 2$, is not absolutely PPT.

Similarly, when $m=3$, inequality \eqref{MNNPPT} becomes:
\begin{equation}\label{3NNPPT28}
\lambda_{2}+\lambda_{3}\geq \lambda_{3n-3}+\lambda_{3n-4}+2\lambda_{3n-5},
\end{equation}
it follows that
\begin{equation*}
  2\lambda_{3n-5}\leq \lambda_{2}-\lambda_{3n-3}+\lambda_{3}-\lambda_{3n-4}.
\end{equation*}
%Since the eigenvalues $\lambda_1, \ldots, \lambda_{3n}$ of $\rho$ are arranged in decreasing order,
It is noted that the third row and column of matrix $\Lambda_{1}^{3}$ exhibit the smallest sum of the absolute values off-diagonal and the largest the absolute values of diagonal element.
This indicates that if \eqref{3NNPPT28} holds, none of the rows of matrix $\Lambda_{1}^{3}$ are diagonally dominant. Therefore, mixed state $\rho$ in the Hilbert space $\mathcal{H}_{3n}$, where $n\geq 3$, is not absolutely PPT.

Lastly, when $m=4$, Theorem \ref{thmMNPPT} aligns with Theorem \ref{thm4NPPT}.
\end{remark}

In summary, we have presented sufficient conditions for absolute separability based on the spectrum and also established sufficient conditions for non-absolute separability. To further illustrate our findings, we provide numerical examples as demonstrations.

\begin{example}
Consider a mixed state $\rho\in \mathcal{H}_{9}=\mathcal{H}_3 \otimes \mathcal{H}_3$ with the following eigenvalues:
$\lambda_9=\lambda_8=\lambda_7=0.0961$, $\lambda_6=\lambda_5=\lambda_4=\lambda_3=0.1111$, and $\lambda_2=\lambda_1=0.1336$.

Based on Lemma \ref{lem1}, the state $\rho$ is absolutely PPT if and only if the following matrices are positive semidefinite:
\begin{equation*}
\Lambda^{3}_{1}=
\begin{bmatrix}
0.1922 & -0.0375 & -0.0225 \\
-0.0375 & 0.1922 & 0 \\
-0.0225 & 0 & 0.2222
\end{bmatrix} \succeq 0,
\end{equation*}
\begin{equation*}
\Lambda^{3}_2=
\begin{bmatrix}
0.1922  & -0.0375 & -0.0375 \\
-0.0375 & 0.2222 & 0 \\
-0.0375 & 0 & 0.2222
\end{bmatrix} \succeq 0 .
\end{equation*}

We can easily verify that
\begin{equation*}
\lambda_9+\lambda_8+\lambda_7 = 0.2883 \ge 0.2672 = \lambda_2+\lambda_1.
%3\times0.0961-2\times0.1336=0.0211>0.
\end{equation*}
Therefore, $\rho$ is absolutely PPT.
In fact, the eigenvalues of $\Lambda^{3}_{1}$ are $\{0.1521, 0.2222, 0.2623\}$ and the eigenvalues of $\Lambda^{3}_{2}$ are $\{0.1510, 0.2122, 0.2435\}$.

On the other hand,  consider a density matrix $\rho\in \mathcal{H}_{9}=\mathcal{H}_3 \otimes \mathcal{H}_3$ with the following eigenvalues:
$\lambda_1=0.6412$, $\lambda_2=0.0923$, $\lambda_3=0.0905$, $\lambda_4=0.0436$,  $\lambda_5=0.0430$, $\lambda_6=0.0311$, $\lambda_7=0.0228$, $\lambda_8=0.0185$, and $\lambda_9=0.0171.$

According to Theorem \ref{thmMNPPT}, we have
$$\lambda_{2}+\lambda_{3} = 0.1828  \ge  0.1613 = \lambda_{6}+\lambda_{5}+2\lambda_{4}.$$
Thus,  $\rho$ is not absolutely PPT. In fact, the eigenvalues of $\Lambda^{3}_{1}$ are $\{-0.5916, 0.0957, 0.6627\}$ and the eigenvalues of $\Lambda^{3}_{2}$ are $\{-0.5849, 0.0970, 0.6714\}$.
By applying Lemma \ref{lem1}, we can obtain the same conclusion.
\end{example}

%\begin{example}
%We consider a density matrix $A\in \mathcal{H}_{3n}=\mathcal{H}_3 \otimes \mathcal{H}_n$ with the following eigenvalues:
%$\lambda_{3n}=\lambda_{3n-1}=\lambda_{3n-2}=\frac{1}{100}-\gamma, \cdots,\lambda_2=\lambda_1=\frac{1}{100}+\frac{3}{2}\gamma.$
%
%According to Lemma \ref{lem1}, matrix $A\in \mathcal{H}_{3n}=\mathcal{H}_3 \otimes \mathcal{H}_n$ is absolutely PPT if and only if the following matrices are positive semi-definite:
%\begin{equation*}
%\Lambda^{3}_{1}=
%\begin{bmatrix}
%2(\frac{1}{100}-\gamma) & -\frac{5}{2}\gamma & -\frac{3}{2}\gamma \\
%-\frac{5}{2}\gamma & 2(\frac{1}{100}-\gamma) & 0 \\
%-\frac{3}{2}\gamma & 0 & \frac{2}{100}
%\end{bmatrix} \geqslant 0,
%\end{equation*}
%\begin{equation*}
%\Lambda^{3}_2=
%\begin{bmatrix}
%2(\frac{1}{100}-\gamma) & -\frac{5}{2}\gamma & -\frac{5}{2}\gamma \\
%-\frac{5}{2}\gamma & \frac{2}{100} & 0 \\
%-\frac{5}{2}\gamma & 0 & \frac{2}{100}
%\end{bmatrix} \geqslant 0 .
%\end{equation*}
%When $\gamma=0.0015$, we have:
%\begin{equation*}
%  3(\frac{1}{100}-\gamma)-2(\frac{1}{100}+\frac{3}{2}\gamma)=0.001>0.
%\end{equation*}
%Therefore, based on Theorem \ref{neccery_mn}, there is
%$$\lambda_{3n}+\lambda_{3n-1}+\lambda_{3n-2}\geq \lambda_2+\lambda_1.$$
%
%In fact, the eigenvalues of $\Lambda^{3}_{1}$ and $\Lambda^{3}_{2}$ are
%$\{0.0129, 0.0190, 0.0221\}$, and  $\{0.0130, 0.0200, 0.0240\}$ respectively.
%Thus, we can assert that $A\in \mathcal{H}_{3n}=\mathcal{H}_3 \otimes \mathcal{H}_n$ is absolutely PPT.
%\end{example}

\begin{example}
Consider a mixed state $\rho \in \mathcal{H}_{16}=\mathcal{H}_4 \otimes \mathcal{H}_4$ with the following eigenvalues:
$\lambda_{16}=\lambda_{15}=0.0475$, $\lambda_{14}=\lambda_{13}=\cdots=\lambda_3=0.0625$, and $\lambda_2=\lambda_1=0.0775$.

According to Theorem \ref{thm4}, for a mixed state $\rho\in \mathcal{H}_{16}=\mathcal{H}_4 \otimes \mathcal{H}_4$ to be absolutely PPT, it is necessary and sufficient for the twelve matrices $\Lambda_{t}, t=1,\dots,12$ to be positive semidefinite. Given that $\lambda_{14}=\lambda_{13}=\cdots=\lambda_3=0.0625$, we can focus on verifying the positive semidefiniteness of each of the following matrices $\Lambda_{t}$ based on Corollary \ref{coronly1}:
\begin{equation*}
\Lambda_{1}=
\begin{bmatrix}
0.0950 & -0.0300 & -0.0150 & 0 \\
-0.0300 & 0.1250 & 0 & 0\\
-0.0150 & 0 & 0.1250 & 0\\
0 & 0  & 0 & 0.1250
\end{bmatrix}.
\end{equation*}

It can be easily verified that
%\begin{equation*}
%  2\times 0.0475+2\times 0.0625-(0.0625+2\times 0.0775)=0.0025>0,
%\end{equation*}
%which implies
$$\lambda_{16}+\lambda_{15}+\lambda_{14}+\lambda_{13} = 0.22 \geq 0.2175 = \lambda_{3}+\lambda_2+\lambda_1.$$
Therefore, $\rho$ is absolutely PPT.
In fact, the eigenvalues of $\Lambda^4_{1}$ are given by
%$$eig(\Lambda^4_{1})=\{0.0733, 0.1250, 0.1250, 0.1467\}.$$
$\{0.0733, 0.1250, 0.1250, 0.1467\}$.
Thus, we obtain the same conclusion.

On the other hand, let's consider a density matrix $\rho \in \mathcal{H}_{16}=\mathcal{H}_4 \otimes \mathcal{H}_4$ with the following eigenvalues:
$\lambda_{1}=0.4894$, $\lambda_{2}=0.0897$, $\lambda_3=0.0812$, $\lambda_4=0.0653$, $\lambda_5=0.0459$, $\lambda_6=0.0449$, $\lambda_7=0.0432$, $\lambda_8=0.0220$,
$\lambda_{9}=\cdots=\lambda_{14} =0.0168$, $\lambda_{15}=0.0154$, and $\lambda_{16}=0.0026$.

According to Theorem \ref{thm4NPPT}, we have
$$\lambda_{3}+\lambda_{5}+\lambda_{6}-(2\lambda_{7}+\lambda_{8}+\lambda_{13}+\lambda_{10})=0.0300>0.$$
Thus,  $\rho$ is not absolutely PPT.
In fact, the eigenvalues of $\Lambda^4_{1}$ are given by
%$$eig(\Lambda^4_{1})=\{-0.4781, 0.0447, 0.0965, 0.4955\}.$$
$\{-0.4781, 0.0447, 0.0965, 0.4955\}$.
From these eigenvalues, we can also conclude that $\rho$ is not absolutely PPT.
\end{example}

The aforementioned examples demonstrate the validity of these sufficient conditions, which can be easily determined. These conditions rely solely on the first few leading and
last few leading eigenvalues of the density matrices, enabling us to efficiently determine whether a bipartite state in $\mathcal{H}_{mn}$ is absolutely separable.

In essence, we can ascertain whether the majority of bipartite states in the Hilbert space $\mathcal{H}_{mn}$ are absolutely separable. It is worth noting that determining the positive semi-definiteness of a series of large-scale parameter matrices in high dimensions is an NP-hard problem. However, our method allows us to quickly determine the separability of most bipartite states without sacrificing accuracy.

\section{Conclusions}
In this paper, we explored separability from the spectrum for qudit-qudits states. We have introduced that a state in the Hilbert space $\mathcal{H}_{4n}$, $n \ge 4$,  is absolutely separable if and only if the twelve matrices of symmetric matricizations of eigenvalues of density matrices are positive semidefinite. However, determining the twelve linear matrix inequalities is a challenging task.. In order to simplify the criterion and make it easy to implement, we present some sufficient conditions for absolutely separable states and not absolutely separable states in $\mathcal{H}_{4 n}$.
%which is based on the fact that all third-order principal submatrices of the twelve matrices given are positive semidefinite.
Moreover, these sufficient conditions that rely only on the first few leading and last few leading eigenvalues can be extended to the general Hilbert space $\mathcal{H}_{mn}$. This not only greatly reduces the complexity of determining absolutely separable states but also improves existing conclusions in any dimension. As applications, we can also derive the distance bounds of eigenvalues and the bounds of purity for general absolutely separable states.

\begin{acknowledgments}
Research of Sze was supported by a HK RGC grant PolyU 15300121 and a PolyU research grant 4-ZZRN. The HK RGC grant also supported the post-doctoral fellowship of Xiong at the Hong Kong Polytechnic University.
\end{acknowledgments}

\appendix
\section{Proof of Theorem \ref{thm4}}
Consider the case where $m=4$ and $n\geq 4$. In this situation, we have $p=4$, $p_{+}=10$, and $p_{-}=6$. Our aim is to determine the set $\sum_{\pm}(4)$.
Since the set $E(\alpha)$ remains unchanged if the elements of the vector $\alpha=(\alpha_1, \alpha_2,\alpha_3, \alpha_4)^{\top}$ are permuted or individually multiplied by $-1$, we can assume without loss of generality that $\alpha_{1}\geq \alpha_{2} \geq \alpha_{3} \geq \alpha_{4}\geq 0$.
There are twelve possible orderings for the numbers $\{\alpha_{i} \alpha_{j}\}_{1\le i,j\le 4}$:
\begin{widetext}
%Let $m=4, n\geq 4$. Then $p=4, p_{+}=10$, and $p_{-}=6$. We shall now determine the set $\sum_{\pm}(4)$. If $\alpha_{1}\geq \alpha_{2} \geq \alpha_{3} \geq  \alpha_{4}\geq 0$, there are such orderings:
$$\alpha_{1}^{2}\geq \alpha_{1} \alpha_{2}\geq \alpha_{1} \alpha_{3}\geq  \alpha_{1}\alpha_{4} \geq  \alpha_{2}^{2}\geq \alpha_{2}\alpha_{3}\geq  \max{(\alpha_{2}\alpha_{4}, \alpha_{3}^{2})}\geq \min{(\alpha_{2}\alpha_{4}, \alpha_{3}^{2})}\geq \alpha_{3}\alpha_{4}\geq \alpha_{4}^{2},$$
$$\alpha_{1}^{2}\geq \alpha_{1} \alpha_{2}\geq \max{(\alpha_{1} \alpha_{3}, \alpha_{2}^{2})}\geq \min{(\alpha_{1} \alpha_{3}, \alpha_{2}^{2})}\geq \alpha_{2}\alpha_{3}\geq \alpha_{3}^{2}\geq \alpha_{1}\alpha_{4}\geq \alpha_{2}\alpha_{4}\geq \alpha_{3}\alpha_{4}\geq \alpha_{4}^{2},$$
$$\alpha_{1}^{2}\geq \alpha_{1} \alpha_{2}\geq \max{(\alpha_{1} \alpha_{3}, \alpha_{2}^{2})}\geq \min{(\alpha_{1} \alpha_{3}, \alpha_{2}^{2})}\geq \alpha_{2}\alpha_{3}\geq \alpha_{1}\alpha_{4}\geq \alpha_{3}^{2}\geq \alpha_{2}\alpha_{4}\geq \alpha_{3}\alpha_{4}\geq \alpha_{4}^{2},$$
$$\alpha_{1}^{2}\geq \alpha_{1} \alpha_{2}\geq \max{(\alpha_{1} \alpha_{3}, \alpha_{2}^{2})}\geq \min{(\alpha_{1} \alpha_{3}, \alpha_{2}^{2})}\geq \alpha_{2}\alpha_{3}\geq \alpha_{1}\alpha_{4}\geq \alpha_{2}\alpha_{4}\geq \alpha_{3}^{2}\geq \alpha_{3}\alpha_{4}\geq \alpha_{4}^{2},$$
$$\alpha_{1}^{2}\geq \alpha_{1} \alpha_{2}\geq \max{(\alpha_{1} \alpha_{3}, \alpha_{2}^{2})}\geq \min{(\alpha_{1} \alpha_{3}, \alpha_{2}^{2})}\geq \alpha_{1}\alpha_{4}\geq \alpha_{2}\alpha_{3}\geq
\alpha_{3}^{2}\geq \alpha_{2}\alpha_{4}\geq \alpha_{3}\alpha_{4}\geq \alpha_{4}^{2},$$
$$\alpha_{1}^{2}\geq \alpha_{1} \alpha_{2}\geq \max{(\alpha_{1} \alpha_{3}, \alpha_{2}^{2})}\geq \min{(\alpha_{1} \alpha_{3}, \alpha_{2}^{2})}\geq \alpha_{1}\alpha_{4}\geq \alpha_{2}\alpha_{3}\geq  \alpha_{2}\alpha_{4}\geq \alpha_{3}^{2}\geq \alpha_{3}\alpha_{4}\geq \alpha_{4}^{2}.$$
Therefore, the set $\sum_{\pm}(4)$ consists of twelve elements, as presented below,
which correspond to the twelve symmetric matricizations $\Lambda_{t}, t=1,\dots,12$ mentioned in Theorem \ref{thm4}.
In short, $\rho$ is absolutely PPT if and only if these twelve matrices $\Lambda_{t}$ are positive semidefinite. The proof is completed.

%\begin{scriptsize}
$$\hat{\Lambda}_{1}=
\begin{bmatrix}
\lambda_{4n}& \lambda_{4n-1}& \lambda_{4n-2}& \lambda_{4n-3}\\
-\lambda_{1}& \lambda_{4n-4}& \lambda_{4n-5}&  \lambda_{4n-6}\\
-\lambda_{2}& -\lambda_{4}& \lambda_{4n-7}&  \lambda_{4n-8}\\
-\lambda_{3}& -\lambda_{5} & -\lambda_{6} & \lambda_{4n-9}\\
\end{bmatrix},
\hat{\Lambda}_{2}=
\begin{bmatrix}
\lambda_{4n}& \lambda_{4n-1}& \lambda_{4n-2}& \lambda_{4n-3}\\
-\lambda_{1}& \lambda_{4n-4}& \lambda_{4n-5}&  \lambda_{4n-7}\\
-\lambda_{2}& -\lambda_{4}& \lambda_{4n-6}&  \lambda_{4n-8}\\
-\lambda_{3}& -\lambda_{5} & -\lambda_{6} & \lambda_{4n-9}\\
\end{bmatrix},
\hat{\Lambda}_{3}=
\begin{bmatrix}
\lambda_{4n}& \lambda_{4n-1}& \lambda_{4n-2}& \lambda_{4n-6}\\
-\lambda_{1}& \lambda_{4n-3}& \lambda_{4n-4}&  \lambda_{4n-7}\\
-\lambda_{2}& -\lambda_{3}& \lambda_{4n-5}&  \lambda_{4n-8}\\
-\lambda_{4}& -\lambda_{5} & -\lambda_{6} & \lambda_{4n-9}\\
\end{bmatrix},
$$
$$
\hat{\Lambda}_{4}=
\begin{bmatrix}
\lambda_{4n}& \lambda_{4n-1}& \lambda_{4n-3}& \lambda_{4n-6}\\
-\lambda_{1}& \lambda_{4n-2}& \lambda_{4n-4}&  \lambda_{4n-7}\\
-\lambda_{2}& -\lambda_{3}& \lambda_{4n-5}&  \lambda_{4n-8}\\
-\lambda_{4}& -\lambda_{5} & -\lambda_{6} & \lambda_{4n-9}\\
\end{bmatrix},
\hat{\Lambda}_{5}=
\begin{bmatrix}
\lambda_{4n}& \lambda_{4n-1}& \lambda_{4n-2}& \lambda_{4n-5}\\
-\lambda_{1}& \lambda_{4n-3}& \lambda_{4n-4}&  \lambda_{4n-7}\\
-\lambda_{2}& -\lambda_{3}& \lambda_{4n-6}&  \lambda_{4n-8}\\
-\lambda_{4}& -\lambda_{5} & -\lambda_{6} & \lambda_{4n-9}\\
\end{bmatrix},
\hat{\Lambda}_{6}=
\begin{bmatrix}
\lambda_{4n}& \lambda_{4n-1}& \lambda_{4n-3}& \lambda_{4n-5}\\
-\lambda_{1}& \lambda_{4n-2}& \lambda_{4n-4}&  \lambda_{4n-7}\\
-\lambda_{2}& -\lambda_{3}& \lambda_{4n-6}&  \lambda_{4n-8}\\
-\lambda_{4}& -\lambda_{5} & -\lambda_{6} & \lambda_{4n-9}\\
\end{bmatrix},
$$
$$
\hat{\Lambda}_{7}=
\begin{bmatrix}
\lambda_{4n}& \lambda_{4n-1}& \lambda_{4n-2}& \lambda_{4n-5}\\
-\lambda_{1}& \lambda_{4n-3}& \lambda_{4n-4}&  \lambda_{4n-6}\\
-\lambda_{2}& -\lambda_{3}& \lambda_{4n-7}&  \lambda_{4n-8}\\
-\lambda_{4}& -\lambda_{5} & -\lambda_{6} & \lambda_{4n-9}\\
\end{bmatrix},
\hat{\Lambda}_{8}=
\begin{bmatrix}
\lambda_{4n}& \lambda_{4n-1}& \lambda_{4n-3}& \lambda_{4n-5}\\
-\lambda_{1}& \lambda_{4n-2}& \lambda_{4n-4}&  \lambda_{4n-6}\\
-\lambda_{2}& -\lambda_{3}& \lambda_{4n-7}&  \lambda_{4n-8}\\
-\lambda_{4}& -\lambda_{5} & -\lambda_{6} & \lambda_{4n-9}\\
\end{bmatrix},
\hat{\Lambda}_{9}=
\begin{bmatrix}
\lambda_{4n}& \lambda_{4n-1}& \lambda_{4n-2}& \lambda_{4n-4}\\
-\lambda_{1}& \lambda_{4n-3}& \lambda_{4n-5}&  \lambda_{4n-7}\\
-\lambda_{2}& -\lambda_{4}& \lambda_{4n-6}&  \lambda_{4n-8}\\
-\lambda_{3}& -\lambda_{5} & -\lambda_{6} & \lambda_{4n-9}\\
\end{bmatrix},
$$
$$
\hat{\Lambda}_{10}=
\begin{bmatrix}
\lambda_{4n}& \lambda_{4n-1}& \lambda_{4n-3}& \lambda_{4n-4}\\
-\lambda_{1}& \lambda_{4n-2}& \lambda_{4n-5}&  \lambda_{4n-7}\\
-\lambda_{2}& -\lambda_{4}& \lambda_{4n-6}&  \lambda_{4n-8}\\
-\lambda_{3}& -\lambda_{5} & -\lambda_{6} & \lambda_{4n-9}\\
\end{bmatrix},
\hat{\Lambda}_{11}=
\begin{bmatrix}
\lambda_{4n}& \lambda_{4n-1}& \lambda_{4n-2}& \lambda_{4n-4}\\
-\lambda_{1}& \lambda_{4n-3}& \lambda_{4n-5}&  \lambda_{4n-6}\\
-\lambda_{2}& -\lambda_{4}& \lambda_{4n-7}&  \lambda_{4n-8}\\
-\lambda_{3}& -\lambda_{5} & -\lambda_{6} & \lambda_{4n-9}\\
\end{bmatrix},
\hat{\Lambda}_{12}=
\begin{bmatrix}
\lambda_{4n}& \lambda_{4n-1}& \lambda_{4n-3}& \lambda_{4n-4}\\
-\lambda_{1}& \lambda_{4n-2}& \lambda_{4n-5}&  \lambda_{4n-6}\\
-\lambda_{2}& -\lambda_{4}& \lambda_{4n-7}&  \lambda_{4n-8}\\
-\lambda_{3}& -\lambda_{5} & -\lambda_{6} & \lambda_{4n-9}\\
\end{bmatrix}.
$$ $\Box$
%\end{scriptsize}
\end{widetext}

\section{Proof of Theorem \ref{thmMNPPT}}

Consider the scenario where $n\geq m\geq 2$. In this case, we have $p=m$, $p_{+}=\frac{m(m+1)}{2}$, and $p_{-}=\frac{m(m-1)}{2}$. Our objective is to determine the set $\sum_{\pm}(m)$. Without loss of generality, we assume that $\alpha_{1}\geq \alpha_{2} \geq \cdots  \geq \alpha_{m}\geq 0$.
There are always such orderings for the numbers $\{\alpha_{i} \alpha_{j}\}$:
\begin{align*}
   & \alpha_{1}^{2}\geq \alpha_{1} \alpha_{2}\geq \alpha_{2}^{2}\geq \alpha_{1} \alpha_{3}\geq \alpha_{2} \alpha_{3} \\
  \geq & \alpha_{3}^{2}\cdots\geq  \alpha_{1}\alpha_{m} \geq   \alpha_{2}\alpha_{m}\geq   \cdots\geq \alpha_{m}^{2},
\end{align*}
%$$\alpha_{1}^{2}\geq \alpha_{1} \alpha_{2}\geq \alpha_{2}^{2}\geq \alpha_{1} \alpha_{3}\geq \alpha_{2} \alpha_{3}\geq \alpha_{3}^{2}\cdots\geq  \alpha_{1}\alpha_{m} \geq   \alpha_{2}\alpha_{m}\geq   \cdots\geq \alpha_{m}^{2},$$
there are corresponding matrix $\hat{\Lambda}^{m}_{k}$ and symmetric matrix $\Lambda^{m}_{k}$ as following:
\begin{widetext}
\begin{scriptsize}
\begin{equation*}
\hat{\Lambda}^{m}_{k}=
\begin{bmatrix}
\lambda_{mn}& \lambda_{mn-1}& \lambda_{mn-3}& \lambda_{mn-6}& \cdots &\lambda_{mn-\frac{m(m-1)}{2}}\\
-\lambda_{1}& \lambda_{mn-2}& \lambda_{mn-4}&  \lambda_{mn-7}& \cdots & \lambda_{mn-\frac{m(m-1)}{2}+1}\\
-\lambda_{2}& -\lambda_{3}& \lambda_{mn-5}&  \lambda_{mn-8}& \cdots & \lambda_{mn-\frac{m(m-1)}{2}+2}\\
-\lambda_{4}& -\lambda_{5} & -\lambda_{6} & \lambda_{mn-9}& \cdots& \lambda_{mn-\frac{m(m-1)}{2}+3}\\
\vdots & \vdots & \vdots & \vdots & \vdots & \vdots\\
-\lambda_{\frac{(m-1)(m-2)}{2}+1} & -\lambda_{\frac{(m-1)(m-2)}{2}+2}& -\lambda_{\frac{(m-1)(m-2)}{2}+3} & -\lambda_{\frac{(m-1)(m-2)}{2}+4 }& \cdots  & \lambda_{mn-(\frac{m(m+1)}{2}-1)}
\end{bmatrix},
\end{equation*}
\end{scriptsize}
and
\begin{tiny}
\begin{equation*}
\Lambda^{m}_{k}=
\begin{bmatrix}
2\lambda_{mn}& \lambda_{mn-1}-\lambda_{1}& %\lambda_{mn-3}-\lambda_{2}& %\lambda_{mn-6}-\lambda_{4}&
\cdots &\lambda_{mn-\frac{m(m-1)}{2}}-\lambda_{\frac{(m-1)(m-2)}{2}+1}\\
\lambda_{mn-1}-\lambda_{1}& 2\lambda_{mn-2}& %\lambda_{mn-4}-\lambda_{3}&  %\lambda_{mn-7}-\lambda_{5}&
\cdots & \lambda_{mn-\frac{m(m-1)}{2}+1}-\lambda_{\frac{(m-1)(m-2)}{2}+2}\\
%\lambda_{mn-3}-\lambda_{2}& \lambda_{mn-4}-\lambda_{3}& 2\lambda_{mn-5}&  %\lambda_{mn-8}-\lambda_{6}&
%\cdots & \lambda_{mn-\frac{m(m-1)}{2}+2}-\lambda_{\frac{(m-1)(m-2)}{2}+3}\\
%\lambda_{mn-6}-\lambda_{4}& \lambda_{mn-7}-\lambda_{5} & \lambda_{mn-8}-\lambda_{6} & 2\lambda_{mn-9}& \cdots& \lambda_{mn-\frac{n(n+1)}{2}+4}\\
\vdots & \vdots & \vdots & %\vdots & %\vdots &
\vdots\\
\lambda_{mn-\frac{m(m-1)}{2}}-\lambda_{\frac{(m-1)(m-2)}{2}+1} & \lambda_{mn-\frac{m(m-1)}{2}+1}-\lambda_{\frac{(m-1)(m-2)}{2}+2}& %\lambda_{mn-\frac{m(m-1)}{2}+2}-\lambda_{\frac{(m-1)(m-2)}{2}+3} & %-\lambda_{\frac{(n-1)(n-2)}{2}+4 }&
\cdots  & 2\lambda_{mn-(\frac{m(m+1)}{2}-1)}
\end{bmatrix}.
\end{equation*}
\end{tiny}
\end{widetext}

It is worth noting that the last row and column of matrix $\Lambda^{m}_{k}$ exhibit the smallest sum of the absolute values off-diagonal  and the largest the absolute values of diagonal element. In other words, if the last row of matrix $\Lambda^{m}_{k}$ is not diagonally dominant, then none of the rows of matrix $\Lambda^{m}_{k}$ are diagonally dominant. Specifically, if
\begin{equation*}
\sum\limits_{i\neq m}\Lambda^{m}_{k}(m,i) \geq \Lambda^{m}_{k}(m,m),
\end{equation*}
then the inequality \eqref{MNNPPT} holds, where $\Lambda^{m}_{k}(m,i)$ is the entry of matrix $\Lambda^{m}_{k}$ located in row $m$ and column $i$.

%Given that %the eigenvalues $\lambda_1, \ldots, \lambda_{mn}$ are arranged in decreasing order and
%the construction of $\Lambda^{m}_{k}$ matrices (where $k$ represents the number of matrices), if none of the rows in the matrices of $\Lambda^{m}_{k}$ are diagonally dominant, then none of the $\Lambda^{m}_{k}$ matrices possess diagonally dominant rows.

%By applying Lemma \ref{lemmaND},
Therefore, we can conclude that the $\Lambda^{m}_{k}$ matrix is not positive semidefinite. Consequently, mixed state $\rho$ in the Hilbert space $\mathcal{H}_{mn}$ (where $m\leq n$) is not absolutely PPT. The proof is completed. $\Box$

\bibliography{apssamp}% Produces the bibliography via BibTeX.

\end{document}